\documentstyle[12pt,prc,aps,epsf]{revtex}
\tightenlines
\begin{document}
\draft


\title{Two Phase Simulation of Ultrarelativistic Nuclear Collisions}  

\author{D.~E.~Kahana$^{2}$, S.~H.~Kahana$^{1}$}
 
\address{$^{1}$Physics Department, Brookhaven National Laboratory\\
   Upton, NY 11973, USA\\
   $^{2}$Physics Department, State University of New York, \\
   Stony Brook, NY 11791, USA}
\date{\today}  
  
\maketitle  
  
\begin{abstract}
A two phase cascade is presented for the treatment of ultra-high energy
ion-ion collisions from $\sqrt{s}=17-200$ GeV. First a high energy, fast
cascade is performed, in which original baryons and freed hard partons, if
any, collide. This stage ignores energy loss from soft (slow) processes.  In
this first version in fact, no hard processes, aside from Drell-Yan
production, are included. The space-time history of the hard cascade is then
used as input to reconstruct the soft energy-loss, which occurs over a
longer time scale. Soft meson production is therefore treated as coherent
over groups of interacting nucleons. Two body data, though, are used
extensively to guide this reinitialisation of the cascade, which satisfies
conservation laws for charge, strangeness and of course, energy-momentum. A
second, low energy, cascade is finally carried out among the products of
the hard cascade. The model selected to describe elementary hadron-hadron
collisions in the soft cascade incorporates generic mesons and baryons, which
are the agents for rescattering. We imagine a constituent quark model
applies, with generic mesons consisting of an excited $q\bar q$ pair, and
generic baryons constructed from three quarks. The chief result is a
reconciliation of the important Drell-Yan measurements, indicating high-mass
lepton pairs are produced {\it as if no energy is lost from the nucleons},
with the apparent success of a purely hadronic, soft cascade in describing
nucleon stopping and meson production in heavy ion experiments at the CERN
SPS.

\end{abstract}   
\bigskip

\section{Introduction}  

Many cascades
\cite{frithjof,werner,wang,genericparton,geiger2,RQMD,URQMD,ARC1,LUCIFERI}
have been constructed to consider relativistic heavy ion collisions. Since
the eventual aim of experiments designed to study such collisions is the
creation of a regime in which the quark-gluon structure of hadronic matter
becomes evident, it is ultimately necessary to include partonic degrees of
freedom in such cascades. However, since at SPS and even at RHIC energies it
is by no means clear that all initial or subsequent hadron-hadron collisions
occur with sufficient transverse momentum to free all partons \cite{eskola},
any complete simulation must deal, at least in part, initially with
collisions of the original nucleons, and finally with collisions of all the
produced hadrons. We will regard the extent to which partons play a role in
the initial stage as an open question. In the course of laying out potential
algorithms for the hadronic sector of the cascade, there arises a very
natural, time-scale driven division between hadronic and partonic sectors.
We present in this work the architecture of a new ultra-high energy cascade,
including its hadronic cross-section inputs, and a detailed description of
the algorithms.

To begin, we wish to discuss the global time scales which divide the cascade
into sectors one could designate as `hard, perturbative QCD, or partonic',
and `soft, non-perturbative QCD, or hadronic'.  This classification has been
discussed often in the literature
\cite{eskola,gottfried,koplik,amueller1,schwimmer,bjorken,dokshitzer}.  For
our purposes here a general outline will suffice. Experimental results
\cite{earlypAdata} on both energy loss and forward pion production in high
energy proton-nucleus scattering were the key motivation for this reasoning.
Energy-loss through meson production at low transverse momentum $p_T$ is in
general a `slow' process. In contrast to this stand `fast' or hard processes,
of which Drell-Yan pair production (DY) \cite{NA3,E772,NA51}, which we will
consider in some detail, is a good example. Characteristic time scales for
these processes can in the first approximation be inferred from the
uncertainty principle. Small momentum transfer meson production takes a long
time $\tau_f \sim p_T^{-1}$, while production of lepton pairs with high mass,
in excess of say $M_{\mu\mu} \sim 4$ GeV takes much less time, $\tau_h \sim
1/M_{\mu\mu}$.  

\begin{figure}
\vbox{\hbox to\hsize{\hfil
\epsfxsize=6.1truein\epsffile[24 85 577 736]{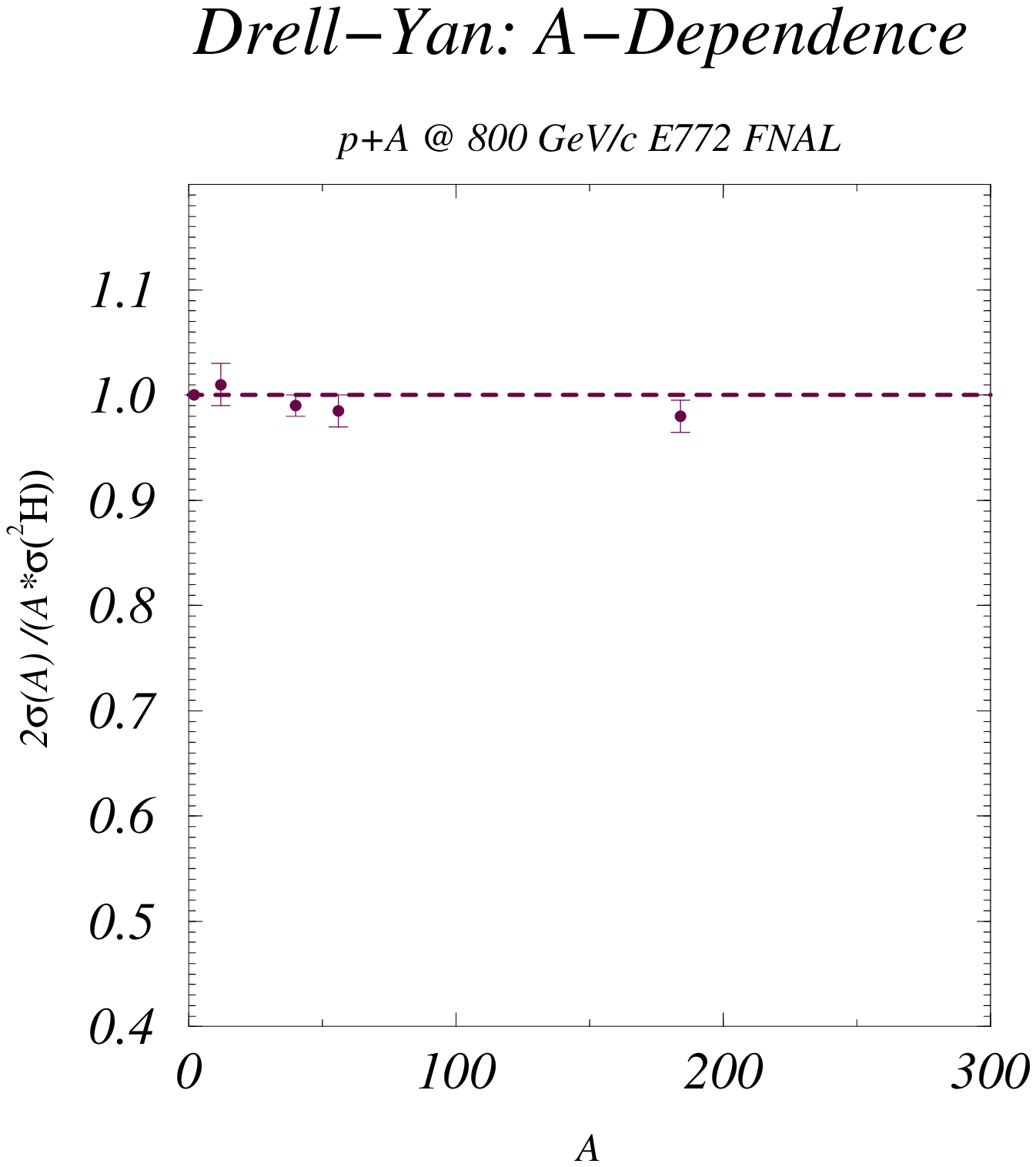}
\hfil}}
\caption[]{The A-dependence of measured Drell-Yan from E772 (FNAL).}
\label{fig:one}
\end{figure}

\begin{figure}
\vbox{\hbox to\hsize{\hfil
\epsfxsize=6.4truein\epsffile[24 85 577 736]{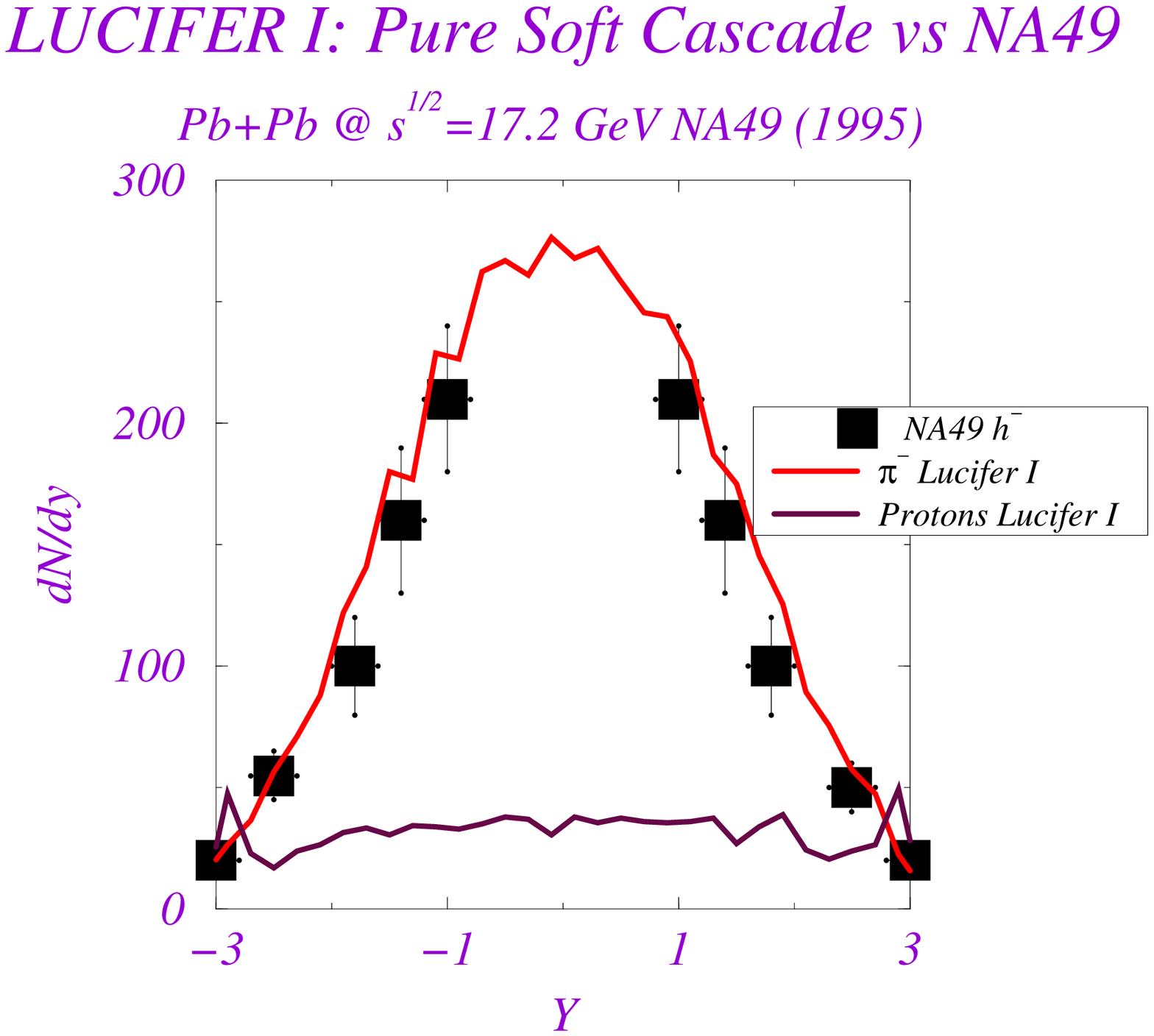}
\hfil}}
\caption[]{Pb+Pb @ 158 GeV/c; Experiment(preliminary NA49, 1995) for Pb+Pb
vs standard cascade theory Lucifer I, from RHIC'96}
\label{fig:two}
\end{figure}

If one examines Drell-Yan data\cite{E772} in high energy pA
collisions, Fig(\ref{fig:one}), the apparent $A$-dependence of the DY
cross-section $\sigma_{DY} \sim A$ suggests that all production of high mass
pairs takes place {\it at the highest energy}, that is, at the energy of the
initial pp collision. The incoming proton simply counts every nucleons that
it hits, and has an equal probability of producing a DY pair in each
collision. This stands in direct opposition to one's na\"ive expectation for
a standard hadronic two body cascade, in which successive collisions of the
leading baryon take place at lower and lower energies, predicting that
$\sigma_{DY}$ should rise more slowly than $A$. Pion numbers found in pA for
the highest Feynman $x$ in similar collisions \cite{earlypAdata} also show
little increase above that expected from pp at the incoming energy. These
results together imply the initiating proton has, in producing DY pairs,
suffered no energy loss in its passage through the nucleus. Nevertheless,
calculations with a purely hadronic cascade \cite{LUCIFERI} describe very
well the large energy loss represented in the inclusive pion spectrum
observed in massive Pb+Pb collisions at SPS energies (see Fig
\ref{fig:two}). It is as if, even though the protons in their initial
interactions inside a nucleus do not lose appreciable energy, they
nevertheless remember full well what number and sort of collisions they have
experienced. The question then is, can the initial energy retention
characteristic of Drell-Yan and eventual energy loss into meson production be
described in a unified dynamic fashion.

It should be noted that the reason for this break down in the pure two body
cascade in ultra-high energy ion-ion collisions is that the soft time scale
$\tau_f$ (for meson production) has become longer than the time scale
$\tau_c$ for the projectile to pass through the target. Therefore soft
energy loss cannot occur before all the hard collisions have taken place.

These apparently contradictory features can in fact be put together in a
single particle or resonance-based multiscattering scheme, simply and in a
rather robust fashion. Gottfried \cite{gottfried} has suggested a mechanism
to describe leading particle behaviour in pA, which we discussed and
implemented in an earlier effort \cite{LUCIFERI} at building a simulation
code, which tracks the time evolution of a highly Lorentz contracted and
excited proton into its final state, including all the mesons expected to be
produced in a free space nucleon-nucleon interaction. These mesons are
inhibited from independently interacting with the target nucleus until they
separate in space from the leading baryon by some minimum hadronic size. This
mechanism ensures that in very high energy pA collisions the leading proton
retains very nearly its incoming energy for most of its passage through the
nucleus. A similar criterion for `re-joining' produced particles was
incorporated by us into a hadronic cascade and this did quite well in the
description of DY production in the FNAL 800 GeV/c pA data \cite{E772}.

\begin{figure}
\vbox{\hbox to\hsize{\hfil
\epsfxsize=6.4truein\epsffile[24 85 577 736]{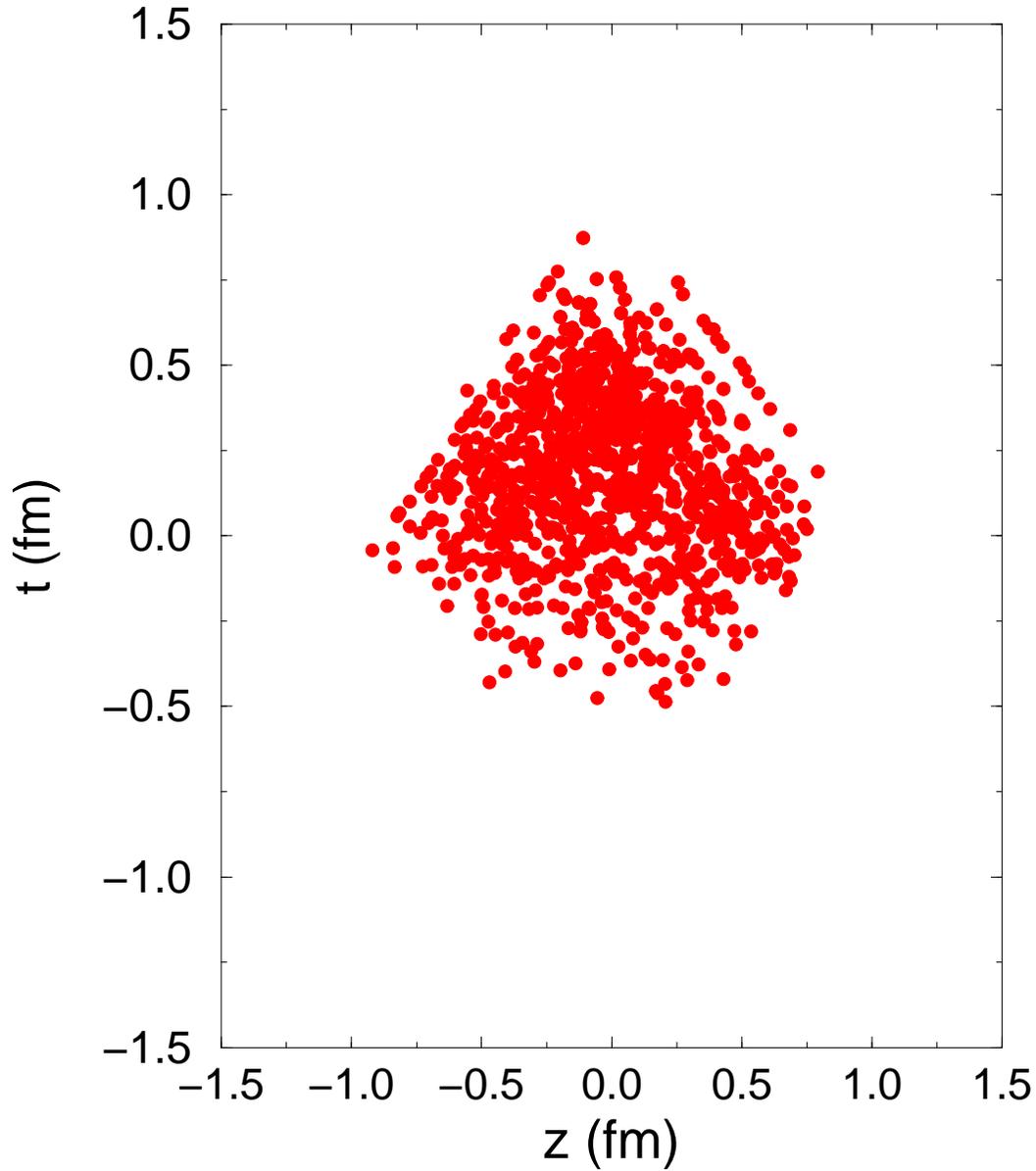}
\hfil}}
\caption[]{Positions of baryons in Pb+Pb after first phase: Fast cascade. The
time and longitudinal envelope indicates where the purely baryonic system
reaches after the two massive nuclei pass through each other at this SPS energy.}
\label{fig:three}
\end{figure}

We have decided here, however, to proceed in what is in the end a simpler
fashion, separating soft and hard processes by time scale, and thus
permitting partonic and hadronic cascading to be put together very naturally
and smoothly, and in a modular fashion. The method simply consists of running
the cascade in two phases, the first a high energy / short time phase in
which collision histories are recorded and fast processes are allowed to
engage. This stage ends at time $\tau_c$, when the projectile and target
nuclei have completely passed through each other.  Then, using the space-time
history of the first phase, a reinitialisation of the cascade is 
performed and a second phase, consisting of a normal soft hadronic cascade
is carried out, at greatly reduced energy and over longer times. Figure
(\ref{fig:three}) shows the final positions of baryons after the first phase,
and the almost light-like paths for the particles engaging in the
initial cascade.

The reinitialization is most critical and is generated from the detailed
space-time history acquired during the initial high energy cascading. The
procedure followed is first to set up groups of nucleons which have mutually
collided in the initial phase of the cascade.  Energy loss can then be
computed for {\it each} nucleon using its known trajectory and collision
history, and the measured inelasticity of the nucleon-nucleon
interaction. Nucleon-nucleon collision data, from the initial energy down,
provide the information on multiplicity distributions, rapidity
distributions, and the correlation of these distributions with energy loss,
which is essential in determining the final baryon four-momenta and hence the
energy and momentum to be deposited on mesons. The method employed is
described in detail below but proceeds as closely as possible to
reconstructing the two-body scatterings experienced by each nucleon
traced. Thus fluctuations, inherent in $NN$ scattering, in the energy loss
and in multiplicity, are mirrored in the reinitialisation.  Generic meson
resonances are then added in, group by group, so that all applicable
conservation laws are obeyed.  The cascade is then re-initialised, and
restarted at the time of the last high energy nucleon-nucleon collision that
occurred in the high energy stage. Generic meson resonances are the principal
rescatterers in the ensuing low-energy cascade together with the still
present, much slowed, baryons.

Partonic casacading at sufficiently high momentum transfer could also have
proceeded during the initial stage, taking place over the short time scale
associated with a high momentum transfer. The energy-momentum taken out of
nucleons by such hard partons would have to be subtracted from that available
for the final production of soft mesons among the `nucleon' groups, but it
would eventually reappear in the meson sector as the hard partons completed
their dynamics and re-hadronised. Indeed the time scale for hadronisation is
comparable to the formation time $\tau_f$ for the soft mesons, so in a first
approximation, the mesons produced by parton dynamics might simply be added
to the reinitialisation before restarting the final soft phase.

One might refer to this two stage cascade as an effective separation into
short and long distance behaviour, a separation created by a factorisation of
the time scales, or equivalently by the momentum transfer involved in each
process. As we mentioned previously, there is an extensive literature on this
subject \cite{gottfried,koplik,amueller1,schwimmer,bjorken,dokshitzer,Sterman}.

\begin{figure}
\vbox{\hbox to\hsize{\hfil
\epsfxsize=6.4truein\epsffile[24 85 577 736]{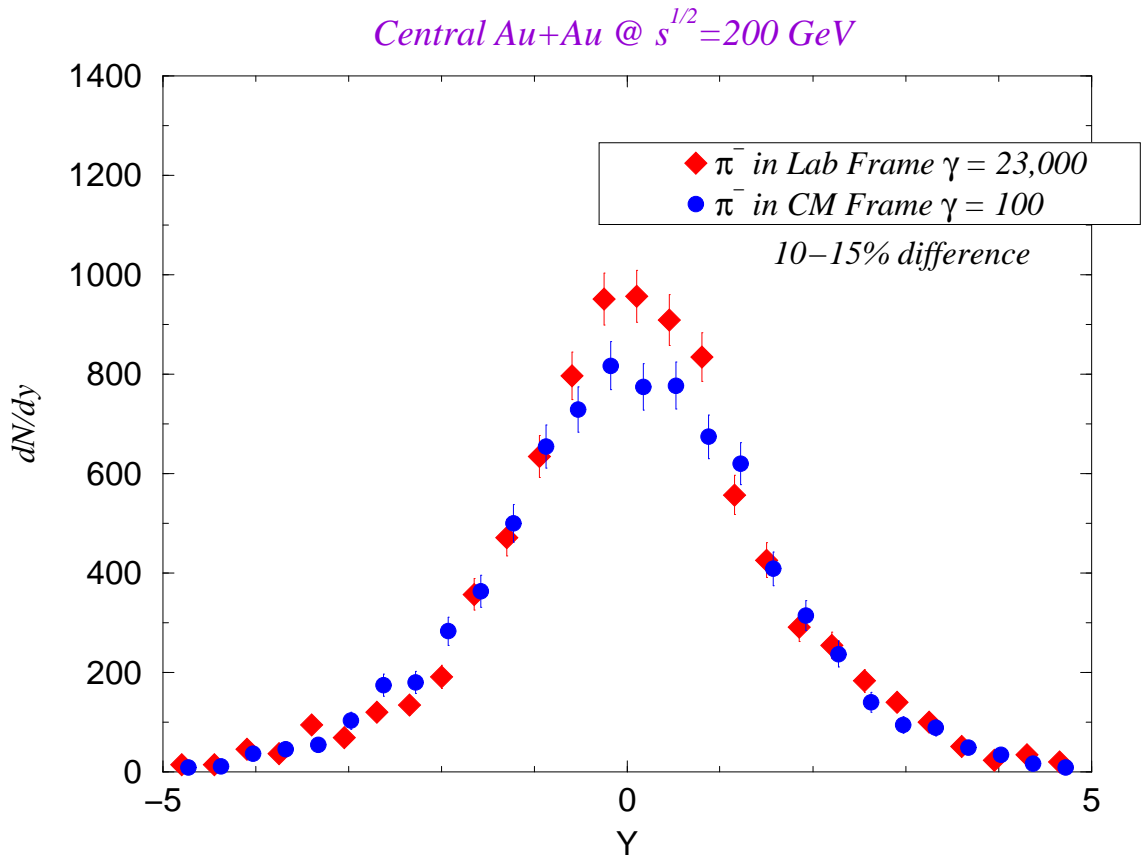}
\hfil}}
\caption[]{A worst case scenario for frame dependence: comparison of spectra 
for a b=0 Au+Au collision at RHIC. The results are from a very preliminary
calcuation with LUCIFER II and do not represent a prediction, but only serves
to indicate the greatly reduced frame dependence. At the SPS the dependence
is within stastical errors.} 
\label{fig:four}
\end{figure}

The two phase approach discussed here has a highly beneficial effect on what
one might fear could be a serious shortcoming of relativistic cascades in
general, that there might be a strong frame dependence. Preliminary
calculations at RHIC energy, $\sqrt{s}=200$ GeV/A, for a central Au+Au
collision, constituting in the present context a worst case scenario, show
that the pion $dN/dy$ differs by no more than 10-15\% between two extreme
cases: a cascade carried out in the global lab frame with $\gamma \sim$
20,000, and a cascade performed in the global center of mass frame with
$\gamma \sim$ 100 (see Figure \ref{fig:four}). The results in the latter
figure must be considered very preliminary for RHIC collisions and represent
no more than a rough prediction for mesons, still less for protons. We intend
here only to illustrate the worst possible extent of the frame dependence to
be expected in our cascade at RHIC energies. This dependence is moderate as
indicated. At SPS energy, the frame dependence is very small, any differences
being negligible and within present statistical errors.

The reason for the minimal frame dependence is simple and systemic: the
initial high energy cascade contains only particles travelling along nearly
light-like paths (see Fig(\ref{fig:three})), and the light cone is, of course,
an invariant surface. Therefore, {\it the re-initialisation does not depend
upon the global frame chosen for the cascade}. Any frame dependence enters
only in the second phase, which involves greatly reduced collision energies
and particle densities. The separation into two stages thus greatly
ameliorates frame dependence.

The second phase involves generic resonances, strange and non-strange, with
the quantum numbers of the $\Delta$, the $N^*$, the $\Sigma$ and the
$\Lambda$ in the baryon sector, and those of the $\rho$, and the $K$ in the
meson sector, with masses in the range $m_B \sim 1$--$2$ GeV for the baryons
and $m_M \sim 0.3$--$1$ GeV for non-strange mesons, and appropriately higher
for strange mesons and baryons. That the non-strange generic meson resonance
mass should be centered near $600-700$ MeV should be no surprise, and is
consistent with our simple picture for these objects, that they comprise a
constituent $q\bar q$ bound pair.

Produced mesons are of course not allowed to interact until a formation time
$\tau_f$ has passed, this is then one {\it real} parameter in the model,
perhaps determinable from pA, or as in earlier work from collisions of light
nuclei \cite{LUCIFERI}. Finally, these resonances decay by pion emission into
observable $\pi$'s, $\rho$'s, $K$'s, hyperons and nucleons, and as many
additional resonances as one cares to include. In this early work we limit
ourselves to $\pi$'s, $\rho$'s and $K$'s. The decay time for the generic
resonances constitutes a second free parameter $\tau_d$. We will need
here, even more than in cascading at the lower energies of the AGS, to use
empirical knowledge of pA collisions and also perhaps light ion data to
predict the most massive ion collisions.

The statement is often made that one's lack of knowledge of
resonance-resonance interactions opens up to cascade models a very large
well of parameters. This is emphatically not the case here. We do not treat
resonance interaction cross-sections as free, but instead employ, as did
Gottfried \cite{gottfried} a universality principle for soft
interactions. Surely, for soft baryon-baryon interactions, most detailed
properties of the resonances are irrelevant, with perhaps only sizes and
reaction thresholds playing any role in determining the nature of the
interaction, which must after all be created by multi-gluon exchange. In fact,
at this point, we also ignore size differences between the resonances, with
the exception of very small objects such as charmonium mesons. This limits
the number of free parameters in the model to a minimum, so far just the two
times $\tau_f$ and $\tau_d$. The important proton-proton and meson-proton
data, over the range of energies at which cascading takes place, are the
primary input and are determined from existing experimental
measurements. Baryon-meson interactions are similarly fixed from known
data. Meson-meson cross-sections are normalised to meson-baryon and
baryon-baryon by appealing to constituent quark model counting.
 
No attempt has been made here to fit the low energy cascade smoothly onto AGS
energies. The cross-sections have been followed sufficiently far down in
energy, however, to ensure that the broad features of spectra at SPS and
certainly RHIC energies are adequately determined within the model.  In this
first presentation, we focus our attention on calibrating and testing the
code on known data at the CERN SPS, and give only one example of a
preliminary calculation at $\sqrt{s}=200$, Fig(4).

The material is laid out as follows. In Section II the fundamental input to
the cascade, the hadron-hadron modelling, is described in detail. Section III
considers the initial high energy phase and then the reinitialisation, both
in nucleus--nucleus collisions. In Section IV, the final phase, a soft
cascade, is described and applied to SPS data. Drell-Yan simulations are
treated in Section V while Section VI presents further results, conclusions
and points out future directions. 

\section{Model for Hadron-Hadron Interactions}

The objective of the cascade approach to calculating nucleus-nucleus
collisions is to proceed from a knowledge of elementary hadron-hadron
collisions to a prediction of the far more complex many body event. This is,
as will become clear, not completely possible in the environment of
ultra-relativistic collisions. Although it may be an acceptable approximation
to omit the small shell-model-like average field at the high energies we
consider, it is not possible to ignore the time structure of the elementary
collisions. Nor can one neglect the nature of the objects initially produced,
which later rescatter during the cascade. At AGS energies \cite{ARC1},
$E_{cm}\sim 1$--$5$ GeV, the introduction of the most evident, lowest lying
resonances, the $\Delta$, the $N^*$ and the $\rho$, was sufficient for a
reasonably accurate picture of the dynamics. The essential time scale for a
many body collision was set by the resonance lifetimes, all $\sim 1.5$
fm/c. Since this scale was in general shorter than the ion-ion collision time
$\tau_c$, that is, the time taken for the colliding nuclei to pass through
each other, most rescattering involved these resonances.

At the considerably higher energies encountered at SPS and RHIC, $\sqrt{s} =
15-200$ GeV, the Lorentz contraction is much more severe. The time $\tau_c$
is now comparable to (SPS), or smaller than (RHIC), both the resonance
lifetimes $\tau_d$ and the meson formation time $\tau_f$. It is therefore
necessary to be guided more by experiment, in particular by pA and the
lighter ion-ion cases, as well as by existing theoretical treatments of
hadron reinteraction in nuclei\cite{gottfried,koplik,Sterman}. At these
higher energies we should of course also allow for the possibility of partons
being freed from hadrons in collisions of high enough transverse momentum. It
is nevertheless interesting to consider first a model restricted
to soft collisions, including however some rare hard processes.

Many approaches \cite{wang,RQMD,URQMD} have been put forward to handle the
reinteraction, including strings \cite{frithjof,werner}, but we prefer to
retain a particle nature for the cascade. The first necessary element then is
a model for the hadron-hadron system, beginning with nucleon-nucleon but
easily extended to meson-nucleon and ultimately applicable to any two body
hadron-hadron collision. The basic processes are elastic scattering and
inelastic production of mesons. Typical samples of the cross-section fits we
use are shown in Fig(\ref{fig:seven}) and in Figure (\ref{fig:eight}) for the
total pp scattering and inclusive $\Lambda$-K production
respectively. Elastic scattering data exists over a wide range of energies
for pp and in this case we fit the basic angular distributions.

We divide inelastic production into the well known categories
\cite{goulianos}: diffractive scattering, referred to as single diffractive
(SD), and non-single diffractive (NSD) scattering \cite{UA5}. A graphic
description of these processes is given in Fig(\ref{fig:nine}). The SD
process leads to a rapidity gap between one of the leading hadrons and the
rest of the produced mesons, and is associated with the triple Pomeron
coupling \cite{pomeron}, while NSD production is attributed to single (or
multiple) Pomeron exchange and results in the observed meson plateau at
mid-rapidity. We have not made a serious effort to include the rapidity
correlations also implied by the Pomeron exchange model, in the belief that
these effects are very likely to be washed out in ion-ion cascading: however,
no serious obstacle exists to doing so. This multiperipheral model of the
soft production then represents the basis for our development, but it must be
supplemented by an intermediate picture which allows us to apply it, not only
to hadron-hadron interactions in free space but also inside a nuclear
environment. The generic mesons depicted in Fig(\ref{fig:nine}), and the
generic baryons, with rather light masses selected in the ranges suggested
above, constitute the basic elements for rescattering in the second-phase of
the cascade. We reemphasize the $q\bar q$ and $qqq$ nature of the generic
resonances, structures easily related to the constituent quark models which
are frequently used for discussing soft QCD physics.

A further crucial element of the hadron-hadron model is the final state meson
multiplicity distribution and its energy dependence. These distributions for
NSD and SD together are assumed to satisy the KNO scaling law \cite{KNO},
despite deviations from scaling noted in CERN UA5 experiments at the SPS
\cite{UA5}. Our theoretical fits to these distributions, over a wide range of
$\sqrt{s}$, are displayed in Fig(\ref{fig:five}) and
Fig(\ref{fig:six}). These figures demonstrate the reasonably high quality of
the fits. The distributions for SD scattering are taken to reproduce known
measurements as discussed by Goulianos\cite{goulianos}.

Our principal phenomenological sources here are the rapidity\cite{goulianos}
and multiplicity\cite{Blobel,UA5,hhisr} information obtainable from the
elementary collisions, generally nucleon-nucleon or pion-nucleon, but also
both proton and pion-nucleus scattering. Under the umbrella of soft SD and
NSD processes we include also strangeness production. Introduction of
truly heavy flavours, however, should in general be done via hard processes.

Guided by our experience in cascading at the lower AGS energies, we introduce
an intermediate step into the SD and NSD production: we imagine these
processes are mediated by a set of intermediate generic hadron
resonances. These soft, Pomeron-mediated, interactions \cite{pomeron,alberi}
involve small $p_T$ but appreciable energy loss, and hence are thought to
proceed on a slow time scale, $\sim 1$ fm/c in the rest frame of the relevant
particles. For generic mesons we include $I=1$ ($\rho$-like) states with
masses in a range between a minimum $300$ MeV and a maximum at or near $1.0$
GeV, and for simplicity a fixed width $\Gamma \sim 125$ MeV. Many of the
results we obtain would equally well follow from the using a single average
mass for the generic mesons near $600$ MeV, in the hadron-hadron model as
well as later on in AB collisions. All resonances of course eventually decay
into observed stable (under strong interactions) mesons and baryons. The low
masses assigned guarantee that $p_T$ remains small in hadronic processes,
even as the resonances decay.

The generic baryons  are either $I=1/2$ ($N$ or $N^*$-like), or $I=3/2$
($\Delta$-like), with masses restricted to the range $1$--$2$ GeV and widths
similar to those for the mesons. In addition we include resonances analogous
to the $\Lambda$, $\Sigma$ and  $K$ at appropriately higher mass due to
the presence of a strange quark.  To reiterate, using a
single, rather low excitation mass for these objects would affect the
modelling very little. Though there is some freedom in the general choice of
these resonance characteristics, all parameters are ultimately fixed by the
known data. None of this two body information is readjusted, in
nucleus-nucleus interactions; the fundamental connection with the two body
measurements remains inviolate. If, for instance, we were to use a somewhat
different mass range for the generic resonances, the constraint that we must
describe the correct elementary multiplicity distributions would lead to a
change in the number of $\pi$'s into which each generic resonance decays. 
These correlated changes persist also in AB collisions making the results 
more stable than one might imagine to changes in parameter choices for
the hadron-hadron model.

In Figure \ref{fig:ten} we display the final fits to pp at the highest
energies relevant to this work \cite{Eisenberg}, $\sqrt{s}\sim 20$ GeV, as
well as for $p\bar p$ at the extreme energy of the Tevatron $\sqrt{s}=1.8$
TeV. The fit to the total cross-section is shown in Figure \ref{fig:seven}.
Similar descriptions and results are obtainable with the pion-nucleon system,
and of course we avail ourselves of these in the ensuing development. The
total multiplicity distributions has been adjusted to account for the
presence of strangeness.

Our model of the two hadron system is very similar in philosophy to the
multiperipheral models\cite{koplik,amueller1,pomeron}. It is clear from
Figure 9 and the development of this section that a central feature, the
rapidity `plateau' for mesons in NSD scattering, has been incorporated
here. Single diffractive clusters with an observed rapidity gap are also
present.  One might further justify the use of intermediate state, generic
mesons, by the oft described pion clustering seen in early
experiments\cite{beusch,berger,quigg}. The generic resonances, as indicated,
later decay into two or three `stable' mesons, including for now $\pi$'s,
$\rho$'s and $K$'s.

The most important features of the soft hadron-hadron system input for the
ion-ion cascade are now set: first the energy loss, second the multiplicity
distribution, and in addition the fluctuations in both of these resulting
from two-body dynamics and geometry.  In particular, the degree of energy
loss in SD and NSD scattering is highly correlated to the width in rapidity
of the final $\pi^-$ distribution in Fig(\ref{fig:ten}). This in turn enables
a surprisingly high energy loss for protons in the rather light S+S system,
in agreement with experiment. More baryons than one might expect end up at
mid-rapidity in central S+S collisions\cite{harris,NA35,NA49}.

We finish this section by restating the universality principle
\cite{ARC1,gottfried} we employ for in-medium resonance-resonance
interactions.  These cross-sections are all, aside from obvious adjustments
in energy threshholds, taken equal to their free space analogues, that is
they are set equal to the appropriate one of $\sigma_{NN}$, $\sigma_{N\pi}$
or $\sigma_{\pi\pi}$. This principle worked well in the cascade ARC, used at
the AGS, and it has the virtue of excluding many adjustable parameters. One
might well expect such soft QCD, multi-gluon exchange interactions to depend
on the sizes of the initial hadrons involved, but given the narrow range of
masses chosen for the generic resonances, size should not be a large
factor. The $c\bar c$ states such as $J/\psi$ with their characteristic small
size are obviously exceptional.

\section{Nucleus-Nucleus Simulation}

One of the puzzles posed above is the apparent contradiction between the
observed linear A-dependence of the production cross-section for massive
Drell-Yan pairs in proton-nucleus and nucleus-nucleus
\cite{NA3,NA35,NA49,NA50,NA51,E772} scattering, and the observed energy
loss\cite{LUCIFERI,NA35,NA49,NA50}. The Drell-Yan process is presumably hard,
rapid and describable by perturbative QCD, while meson production is mainly
soft, slow and non-perturbative in nature. We now try to exploit this
difference in time scales, a difference which is itself a function of
collision energy, to create a global cascade incorporating both hard and soft
processes. To do this we separate the simulation into two phases, connected
by an overall reinitialisation. The first phase, a fast cascade, is
restricted to recording the space-time geometry of the nucleus-nucleus
collision, as a series of nucleon-nucleon collisions, while permitting energy
loss to occur only from sufficiently rapid processes. In this work only the
rare processes of Drell-Yan and charmonium production are used as examples of
hard processes, and more general production of hard partons is left for
future investigation. Such an approach will be sufficient to answer the most
interesting present question: Can one rule out a purely hadronic explanation
of charmonium suppression seen at the SPS?

\subsection{The High Energy Cascade}

The geometrical picture of the initial high-energy cascade on incoming
nucleons in both target and projectile has already been shown in
Fig(\ref{fig:three}). The method used here is straightforward and indeed
closely resembles, in outline, an eikonal or Glauber approximation. However,
the random, fluctuating nature of a cascade is retained. In each event, the
collision between projectile and target is initiated with the correct energy
and a nucleon configuration randomly selected within the allowable nuclear
geometries. The two-body cross-sections, evaluated at the incoming energy and
again employing Monte-Carlo, are used to trace out the collisions and
particle trajectories in phase space. This history is later used to fix the
nucleon positions in space-time as well as to calculate the energy-momentum
losses.  Should a hard collision intervene in the initial phase, its
associated energy loss and production characteristics would also be recorded,
and the energy lost would become unavailable for soft meson production. This
energy, if transferred to a perturbatively interacting parton, could
nevertheless reappear in the final soft cascade through hadronisation of the
parton. 

The time evolution of the high energy phase is completely determined by
normal cascading with nucleons proceeding along straight lines between
collisions. However, in the first stage only elastic collisions are allowed
between nucleons (except for the rare production of a charmonium or a
Drell-Yan pair), and the $p_T$ is first fixed at zero.  This of course leads
to the light-cone-like structure seen in Fig(\ref{fig:three}), which as we
pointed out, hopefully limits to an acceptable extent the frame dependence
of the nuclear collision.  One might speculate that, at least to a first
approximation, ignoring the soft energy loss retains within each nucleon a
coherence which is only destroyed much later.

\subsection{Reinitialisation}

We present enough of the structure of the cascade code to illustrate the
information recorded and later reused. The main global data structures are a
particle list and a collision list. As collisions occur they are removed from
the collision list, and the particle list is rearranged, removing any
particles which have collided from the active portion of the list and adding
new particles produced in the collision. In the initial, hard cascade, the
only produced particles are nucleons or possibly charmonia and Drell-Yan
pairs. Pointers are maintained so that after the initial hard cascade is
complete (after the last high energy nucleon-nucleon collision is removed
from the list), the final baryons may be easily traced back to the initial
ones: this is possible since fermion lines must be continuous from the
initial to the final state. In this way, the entire space-time history of the
cascade is kept available as a guide to reinitialisation.

During reinitialisation a further data structure is introduced: the group
list. We construct groups of nucleons in a manner described fully below. For
the purpose of calculating soft energy loss nucleons in these groups are
considered to have interacted coherently with one another.  Into these groups
we place, particle by particle, mesons carrying the lost energy. All
conservation laws are enforced at the group level.

The group structure is virtually dictated by pA collisions where, given the
finite range of the strong interactions and the relativistic $\gamma$'s at
the SPS and at RHIC, the incoming proton collides mainly with a row of target
nucleons in its direct path. A simple, symmetric, generalisation of this
group structure is used for the AA case.  Finally, the energy-momentum lost
is transferred onto the generic mesons, whose multiplicity is fixed by the
record of nucleon-nucleon collisions.  There is only limited freedom in
assigning the groups. Quantitatively, the possible alternatives make little
difference to the results, not least of all because the second phase cascade
takes place at greatly reduced energies, comparable in both SPS and higher
energy RHIC collisions. In fact, the model resembles an extension of the
wounded nucleon model\cite{woundednucleon}. We have essentially labelled each
incoming nucleon with its multifaceted history. One might refer to this as a
`painted' or programmed nucleon model.

The group selection procedure we describe here is essentially topological in
nature. At the completion of the fast cascade one selects, as a seed for the
first group, that nucleon which underwent the maximal number of collisions in
the hard cascade. In a first pass, one then adds, to the group containing
this seed nucleon, all those nucleons with which it collided.  Since we
consider only the highest energy collisions in the fast cascade, these
colliders are all going (in an equal velocity frame) in a direction
`opposite' to the initial seed nucleon. If the seed nucleon originates from
the target, then the colliders come from the projectile, and {\it vice
versa}.  Clearly, in this way, we generally begin with nucleons near the
center of the collision region, where the longest sections of target and
projectile passed through each other.

\begin{figure}
\vbox{\hbox to\hsize{\hfil
\epsfxsize=6.1truein\epsffile[82 190 520 439]{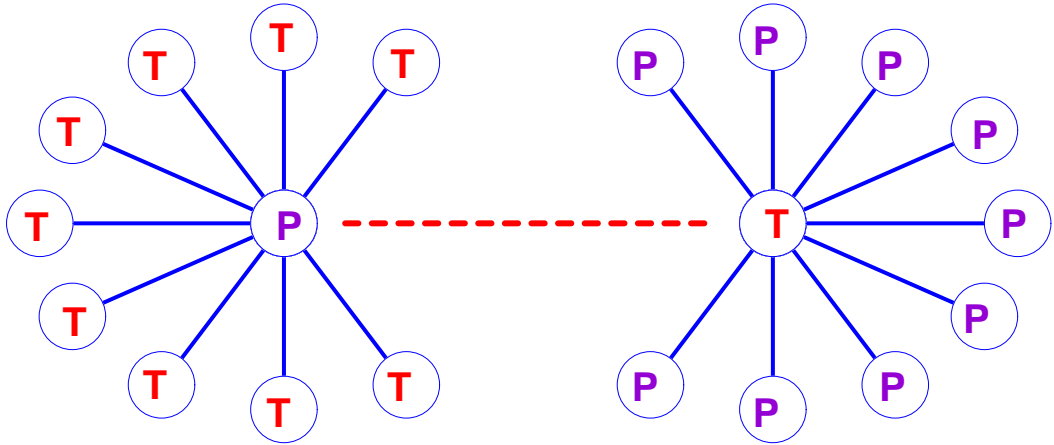}
\hfil}}
\caption[]{A schematic diagram of the structure of the groups used in
reinitialisation of the cascade. Circles inscribed with P stand for
projectile nucleons, circles with T stand for target nucleons. Lines
connecting nucleons indicate that they have collided in the initial
cascade. A group is constructed by first finding a seed nucleon with a
maximal number of collisions in the hard cascade, here represented by one of
the two circles connected by a dotted line. For example, the seed might be
the projectile nucleon at the left of the dotted line.  All target nucleons
which collided with the seed would be added to the group. In a general
nucleus-nucleus collision there would be one among all those target nucleons
added which suffered the largest number of collisions (here drawn at the
right of the dotted line). This nucleon would in turn be used as a second
seed, to symmetrise the group. All projectile nucleons which collided with
the secondary seed would also be added to the group. In an actual example the
groups could of course be larger or smaller than depicted here, and possibly
less symmetric, depending on the relative sizes of the projectile and target,
and the particular geometry of the collision.}
\label{fig:twenty}
\end{figure}

\begin{figure}
\vbox{\hbox to\hsize{\hfil
\epsfxsize=6.4truein\epsffile[24 85 577 736]{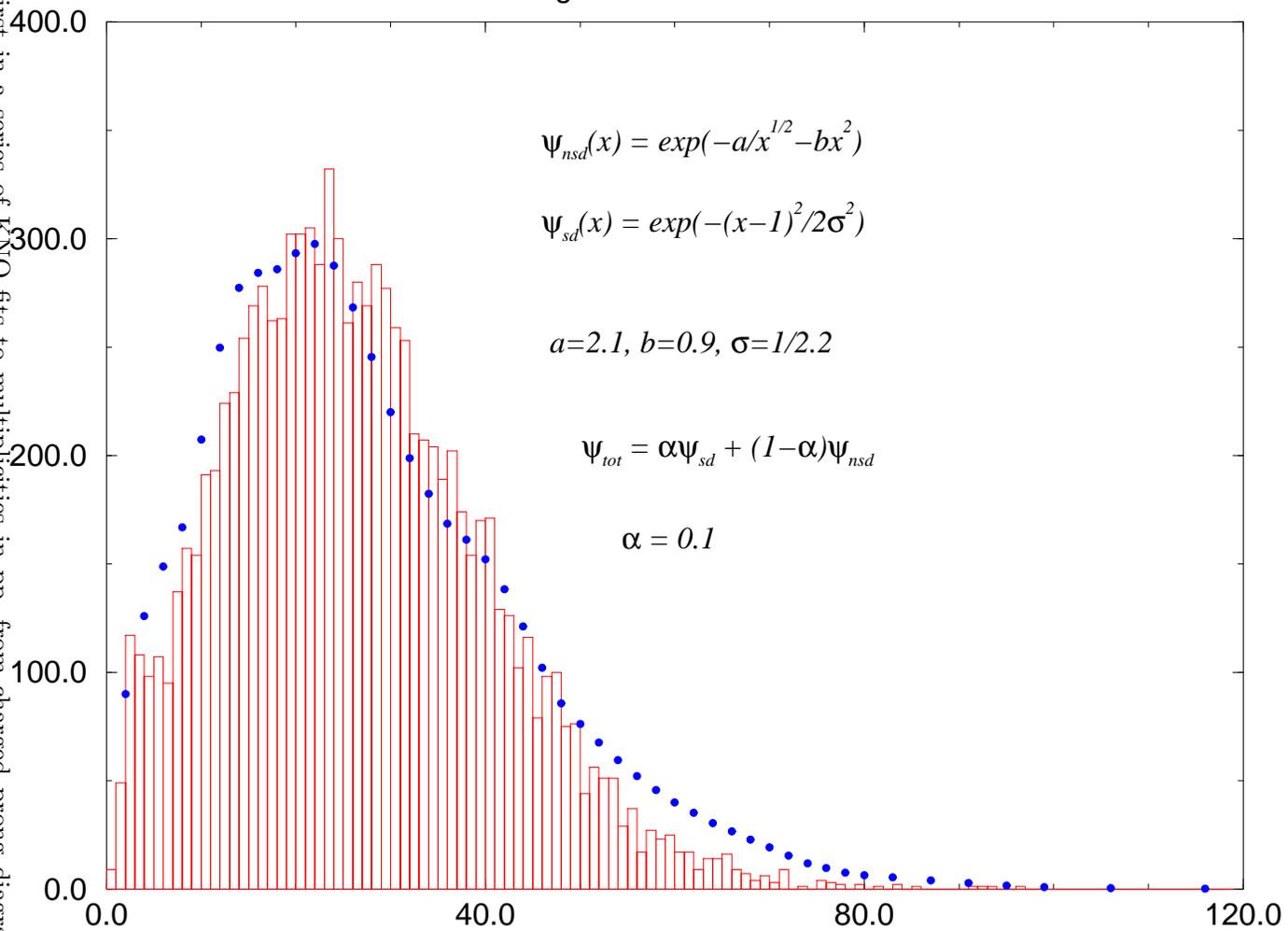}
\hfil}}
\caption[]{First in a series of KNO fits to multiplicities in pp, from
charged prong diagrams at several energies. This depicts the $\sqrt{s}=546$
GeV fit and contains the particular choice of KNO function and parameters
used.}
\label{fig:five}
\end{figure}

\begin{figure}
\vbox{\hbox to\hsize{\hfil
\epsfxsize=6.4truein\epsffile[24 85 577 736]{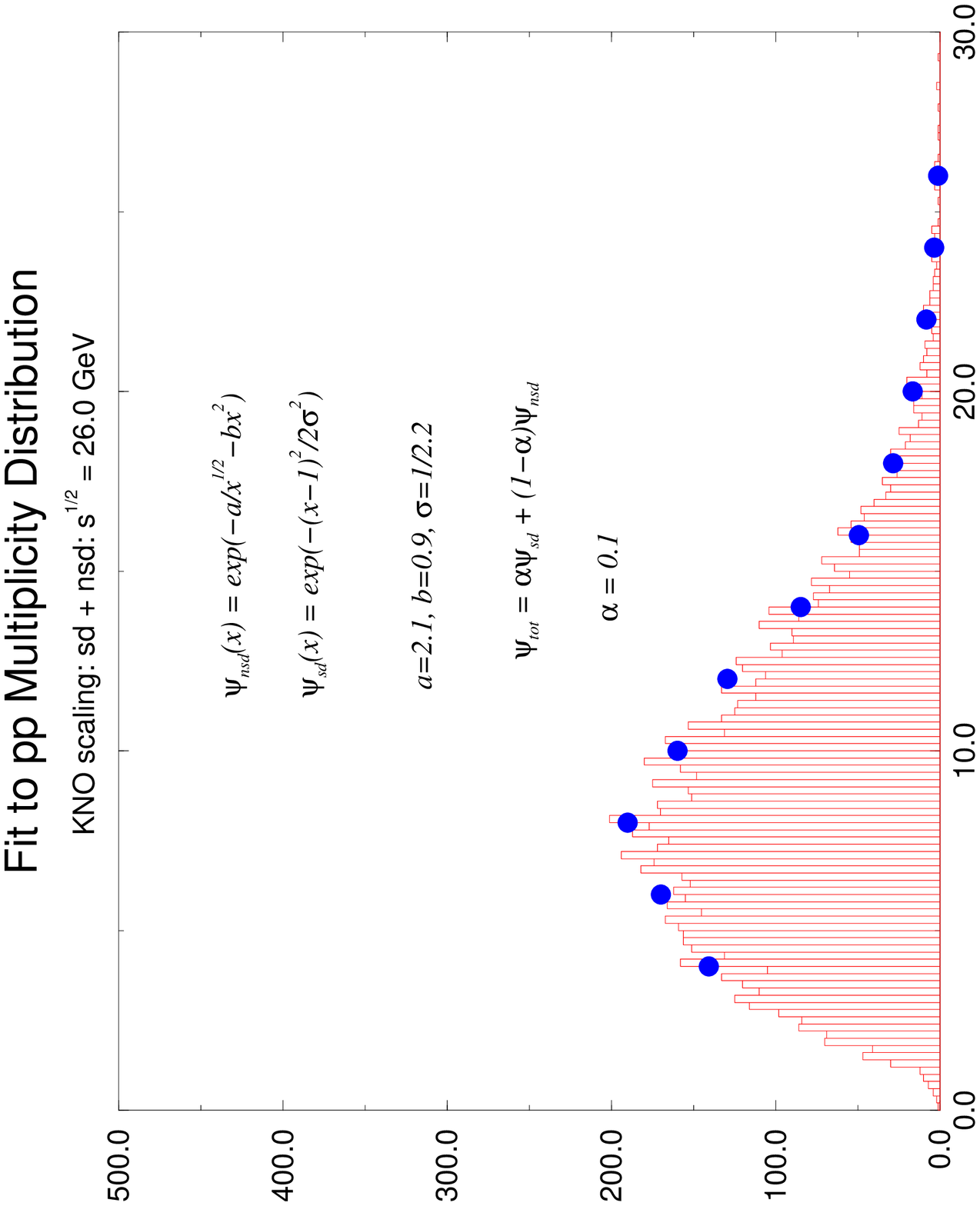}
\hfil}}
\caption[]{Like Figure 5 but for $\sqrt{s} \sim 20$ GeV.}
\label{fig:six}
\end{figure}

\begin{figure}
\vbox{\hbox to\hsize{\hfil
\epsfxsize=6.4truein\epsffile[24 85 577 736]{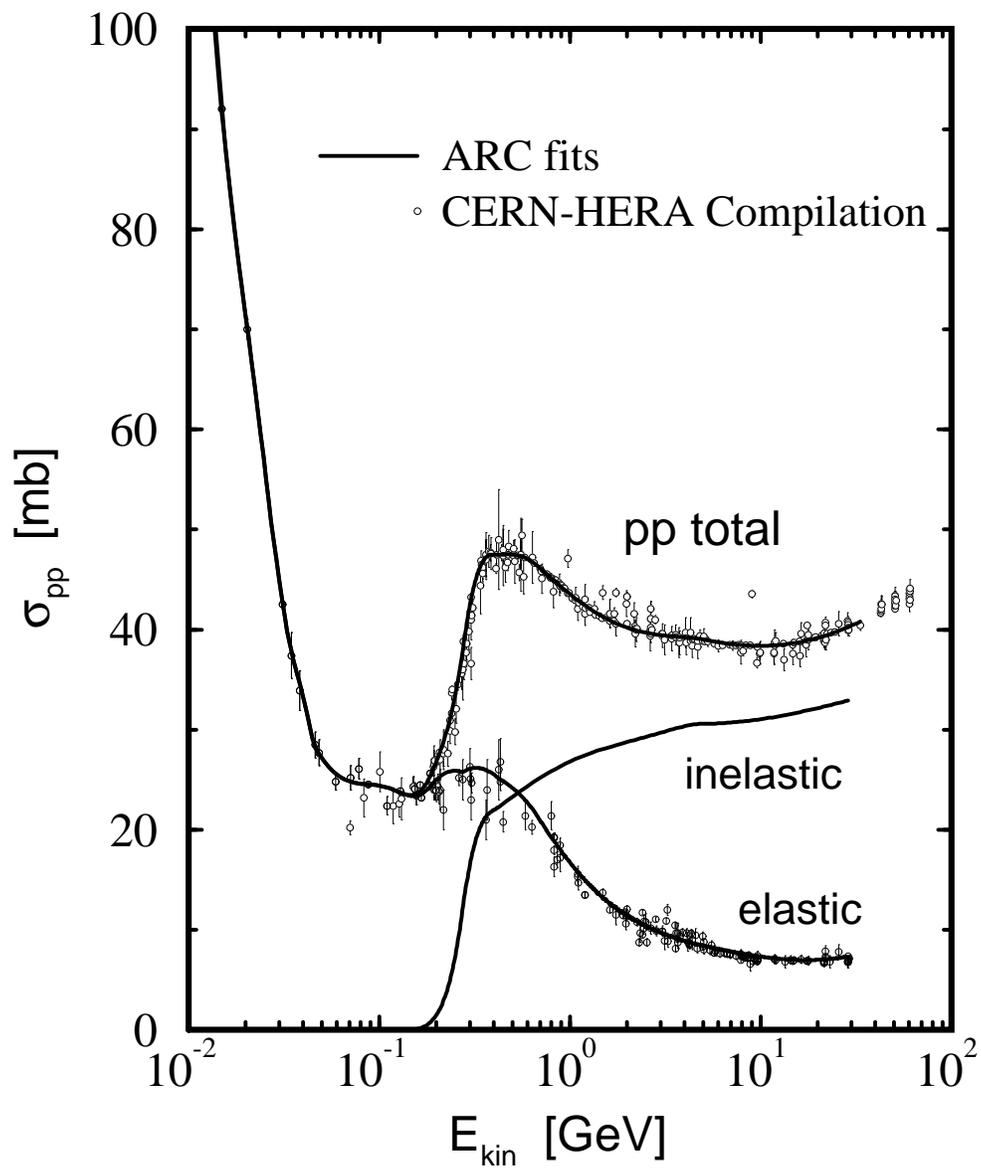}
\hfil}}
\caption[]{Total Cross-section for pp. Experimental points for pp
extends well beyond the CERN SPS  energies, but above one must use
$p\bar p$ measurements. We show only the calculated cross-section beyond
the actual pp measurements}
\label{fig:seven}
\end{figure}

\begin{figure}
\vbox{\hbox to\hsize{\hfil
\epsfxsize=6.4truein\epsffile[24 85 577 736]{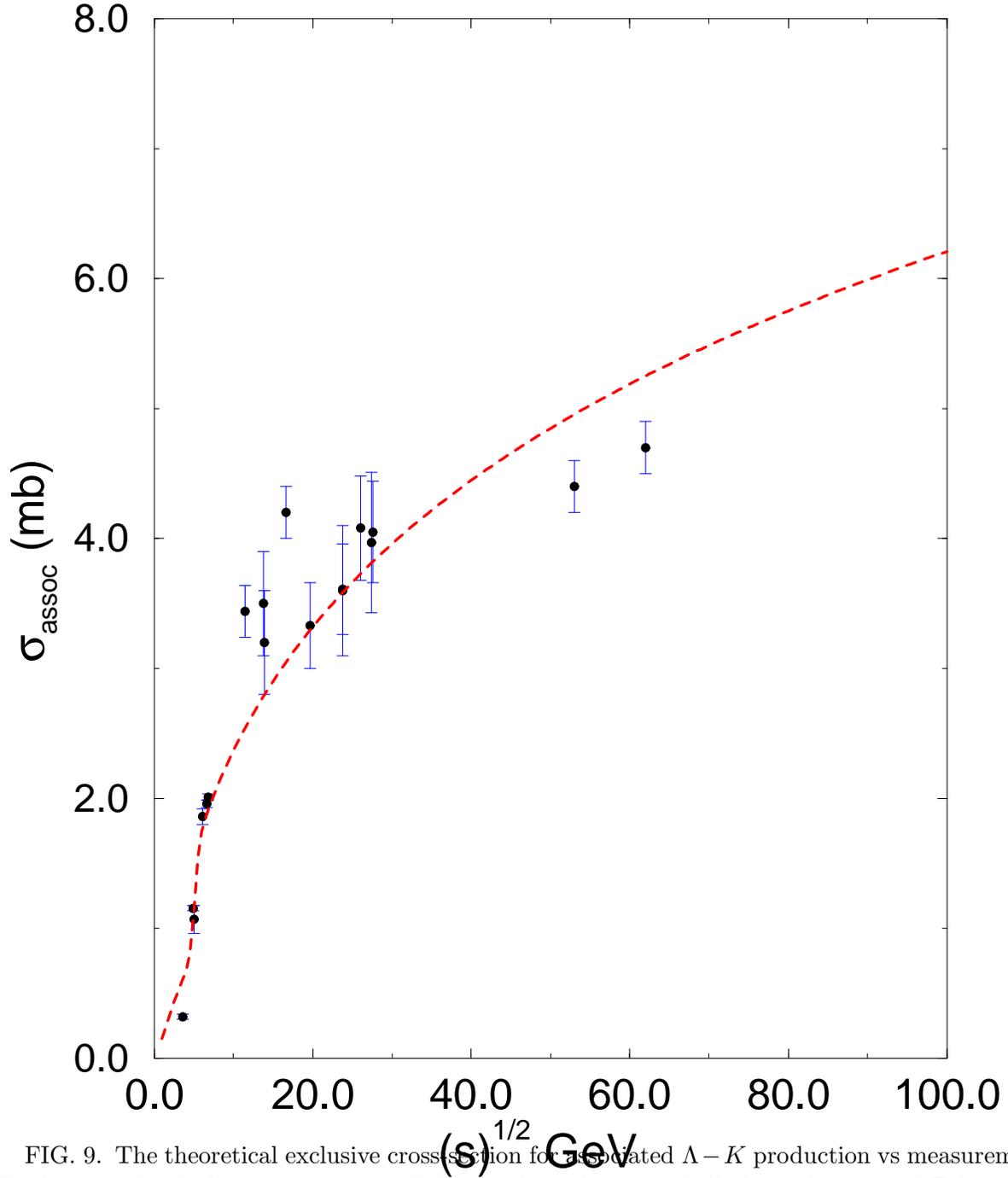}
 \hfil}}
\caption[]{The theoretical exclusive cross-section for associated 
$\Lambda-K$ production vs measurement. In the actual calculations we assumed
the total production of all charged states of $\Sigma$ hyperons equalled that
into $\Lambda$.}
\label{fig:eight}
\end{figure}

\begin{figure}
\vbox{\hbox to\hsize{\hfil
\epsfxsize=6.4truein\epsffile[24 85 577 736]{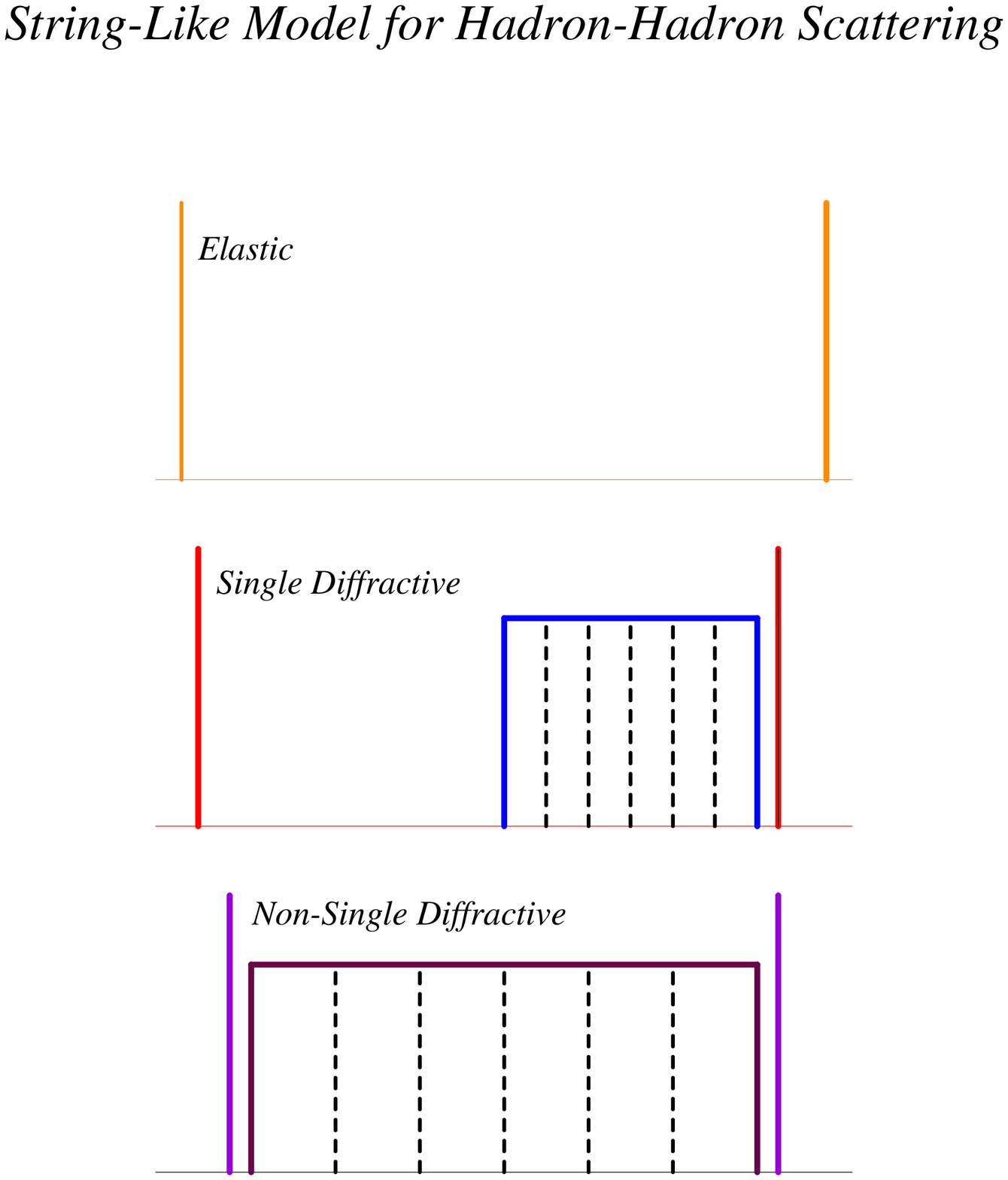}
\hfil}}
\caption[]{Shown are graphic representations of the elements of the model
for the elementary hadron-hadron collision: elastic, single
diffractive (SD) and non-single diffractive (NSD). The meson groups
introduced in both SD (with a rapidity gap) and NSD have a string-like
character but are already sub-divided into generic resonances. It is customary to
associate SD scattering with the three Pomeron coupling and NSD scattering
with multi-Pomeron exchange.} 
\label{fig:nine}
\end{figure}

\begin{figure}
\vbox{\hbox to\hsize{\hfil
\epsfxsize=6.4truein\epsffile[24 85 577 736]{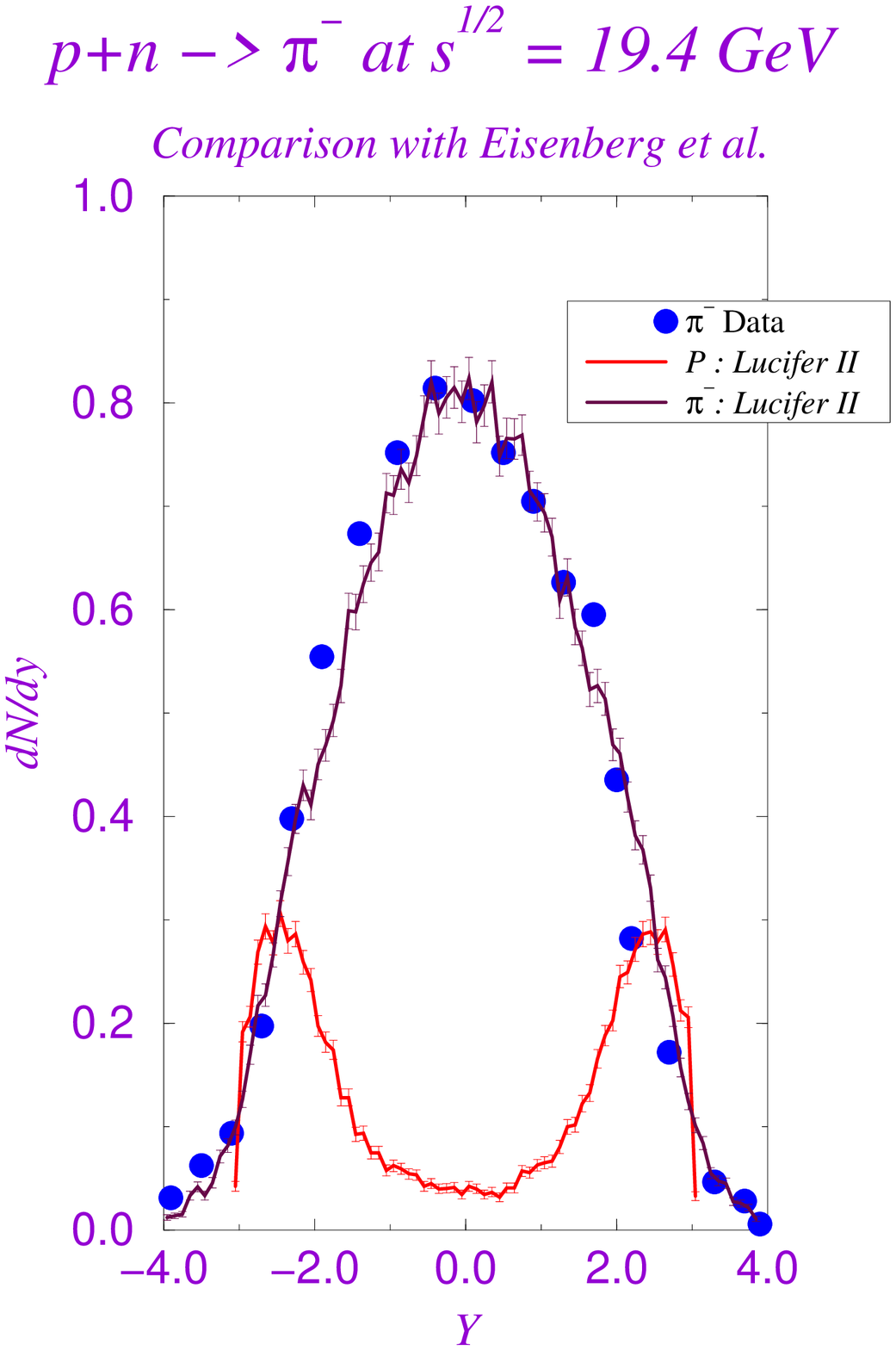}
\hfil}}
\caption[]{The overall fit to the inclusive rapidity spectra for $p$  and
$\pi^-$ from $pn$ scattering at $\sqrt{s}=19.4$, performed by Eisenberg et al.
at FNAL}
\label{fig:ten}
\end{figure}

Although geometry has been carefully left out of this procedure, it plays its
accustomed role, since colliders are separated at most by one cross-section
distance. A second pass is then made to augment and more importantly, to
symmetrise the groups, kinematically speaking, between projectile and
target. The collider nucleons added in the first pass are again ordered by
the number of collisions they suffered, and the maximal nucleon then has its
`opposite' colliders included. The nucleons so included have all interacted
with each other and are separated transversely by at most two cross-section
distances. This completes the first group. A schematic representation of a
group constructed in this way is given in Fig(\ref{fig:twenty}).

One then proceeds similarly for the next group, and so on, until all possible
nucleons have been assigned to groups. At this point there may yet remain
some `orphaned' interacting nucleons, which have interacted but have not
been assigned to any group, and possibly also some spectator nucleons, which
have not had any collisions at all. The `orphans' are assigned to groups at
random, by searching the existing groups for a nucleon with which they have
collided. Finally, non-interacting particles are left individually as
spectators.

This method coincides with the natural definition of an interacting group in
pA collisions. It also leads to groups of reasonable size in both very
massive (Pb+Pb) and lighter (S+S) ion collisions. For example, there are some
46 groups produced in very central Pb+Pb at the SPS with the largest group
containing 25 or so nucleons. The groups generally decrease in size, as one
progresses towards the edges of the interacting region. In a typical central
Pb+Pb event we find as many as ten spectators, whose history is then
self-evident. In central S+S there are of course fewer and smaller
groups. The rest of the calculations in this section are carried out in the
individual group rest frames, after appropriate boosts from the preselected
global frame. Finally, these results are boosted back to the global frame to
and used to initialise the final low energy cascade.

The energy-momentum loss for each nucleon is tied to its collision history,
by reconstructing for each collision it had, using the basic elementary NN
model of Section II, the specific energy loss and the type and multiplicity
of mesons produced. For example, strangeness production, both associated
$\Lambda-K$ and $K-\bar K$, is allowed and is significant at the SPS. A
maximum rapidity for the generic mesons is assigned, again related to that
found in the basic pp interaction, but the final $y$ distribution of mesons
within a group is then mostly dictated by energy-momentum conservation. 
Transverse momentum is added for the baryons assuming
a random walk occurs, with step size $\delta p_T$ taken from basic pp
collisions. The total meson transverse momentum must cancel that in
the baryons. A simple Gaussian parameterisation is used for the meson
phase space in $p_T$. Particular problems of course occur at the edges of
phase space in rare events and these were solved on a case by case basis.

We now describe the method of tracing the collision history of a particular
baryon. The total number of collisions that a given baryon suffered in the
hard cascade is already known: the first of these naturally takes place at
the maximum initial energy. The ratio of elastic to inelastic scattering, as
well as the multiplicity, and type, of mesons created in a nucleon-nucleon
collision follow from our KNO scaling distributions (Figures 5-6) and
cross-section fits. The baryon, in each successive collision, is given a
choice of all possibilities, beginning with elastic or inelastic scattering,
the latter divided into SD, NSD and a small absorptive probability for high
collision number. If an inelastic collision occurs, energy is removed from
the baryon.

These choices are carried out in an effective two-body frame, determined by
searching through the group for the highest energy remaining two-nucleon
collision energy involving the nucleon under consideration. The average
energy loss for baryons, is calculated from the input pp interaction and
multiplicity distributions. At high energies there is a rough rule of thumb
in pp collisions, stating that approximately one-half the energy is lost in
the center of mass frame. This fixed downward drift in rapidity $\Delta y\sim
0.6-0.7$ cannot continue to arbitrarily low energy collisions, but must be
severely limited when low $y$ is achieved.  This is accomplished by adjusting
a Gaussian parametrisation to produce a scaled drift in rapidity, with
gradually decreasing $\Delta y$ at large collision numbers. 

The end result in a given group is a set of baryons with fixed 4-momenta,
positions and also a set of mesons, of specific charges and flavours. There
remains the problem of depositing the implied longitudinal energy-momentum
lost from the baryons onto the generic mesons. If life were simple this would
be a single procedure. In fact to satisfy the final phase space requirement,
energy-momentum conservation in the group, one must allow for some further
readjustment of the baryon momenta with due attention payed to very high
multiplicity cases. In such a case, where the initial attempt to distribute
the energy-momentum among the mesons fails because not enough energy is
available for the mesons, baryon momenta already assigned may be
rescaled. Multiplicity and $p_T$ may also be readjusted in extreme cases to
yield the desired solution.

\begin{figure}
\vbox{\hbox to\hsize{\hfil
\epsfxsize=6.1truein\epsffile[24 59 562 736]{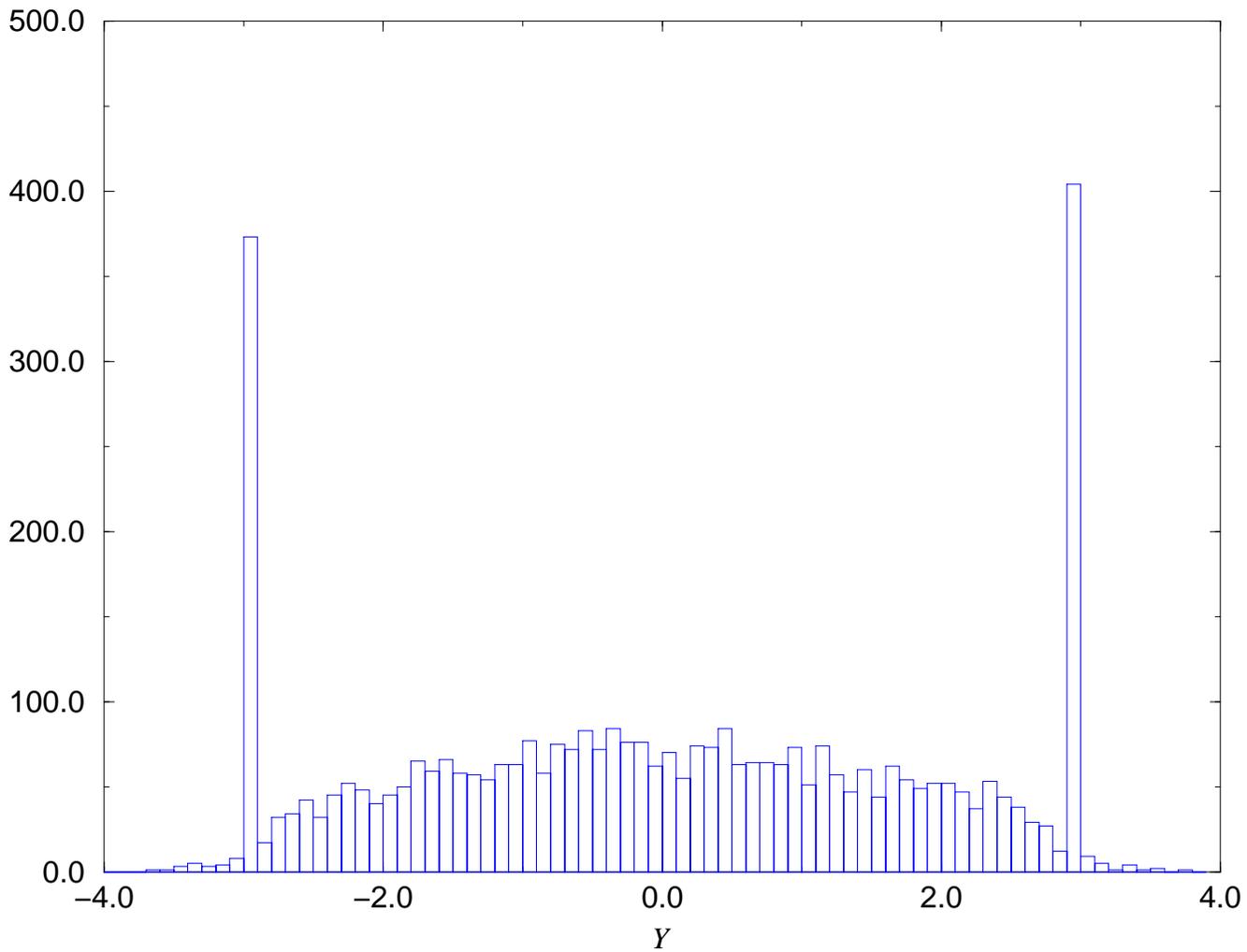}
\hfil}}
\caption[]{Reinitialized rapidity spectrum for baryons in Pb+Pb, i.~e.~just
before the start of the second stage, low energy cascade.}
\label{fig:eleven}
\end{figure}

\begin{figure}
\vbox{\hbox to\hsize{\hfil
\epsfxsize=6.1truein\epsffile[24 59 562 736]{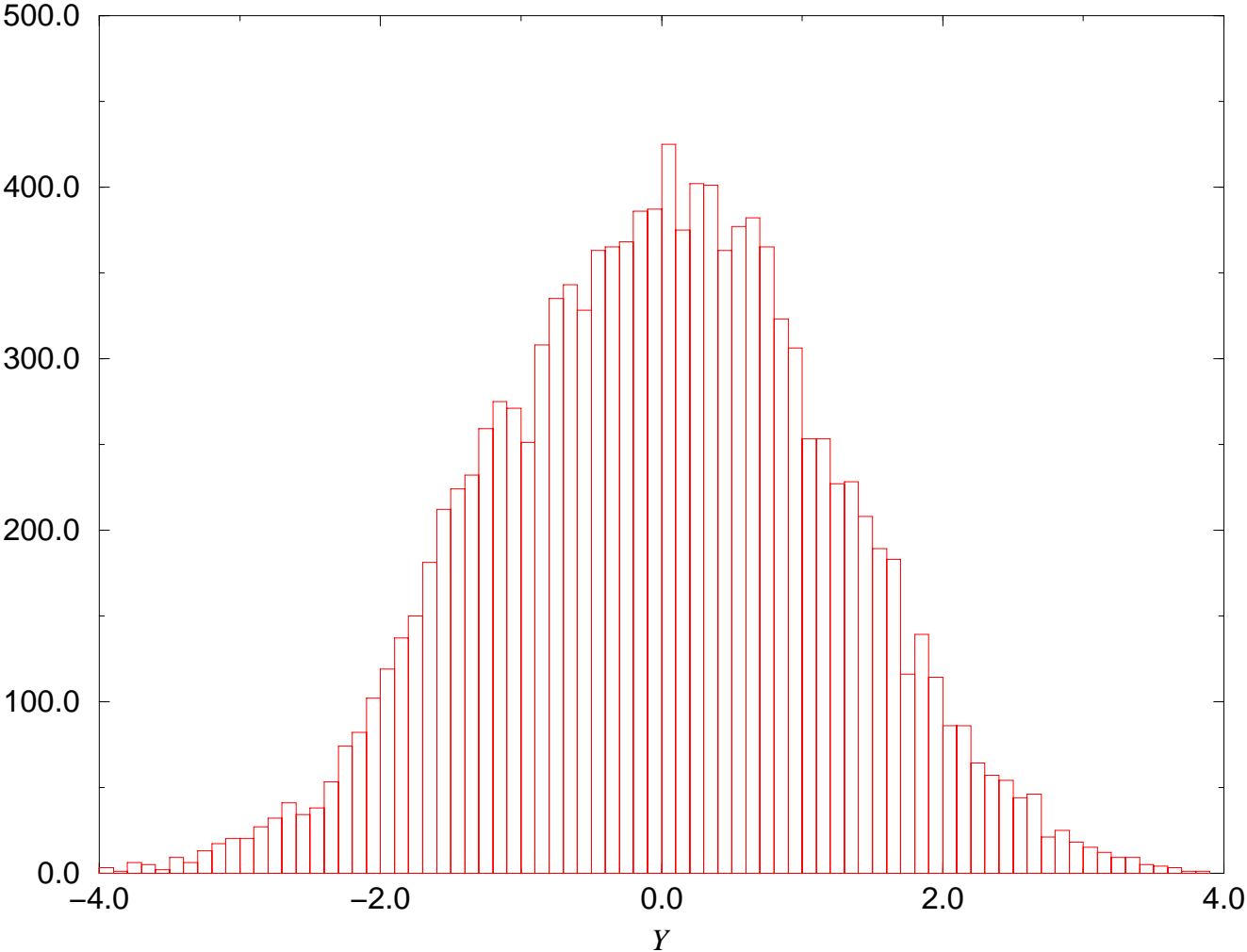}
\hfil}}
\caption[]{Reinitialized rapidity spectrum for generic mesons in Pb+Pb, i.~e.~just
before the start of the second stage, low energy cascade.}
\label{fig:twelve}
\end{figure}

Finally, the mesons and baryons in a group are transformed back to the
preselected global cascade frame. Other alternatives we pursued for
constructing groups and treating phase space, altered the quantitative
results rather little, once sensible solutions were found, so we are
reasonably confident the model is stable. The intricacies of phase space
selection are very strongly dictated by overall energy-momentum conservation,
once the final multiplicity and average $p_T$ is known, and thus the
reinitialisation is again quite robust.  Guidance from the elementary
hadron-hadron data is clearly vital.  Fig(\ref{fig:eleven}) and
Fig(\ref{fig:twelve}) display reinitialised rapidity spectra for both meson
and baryon generic resonances.

\begin{figure}
\vbox{\hbox to\hsize{\hfil
\epsfxsize=6.1truein\epsffile[24 59 562 736]{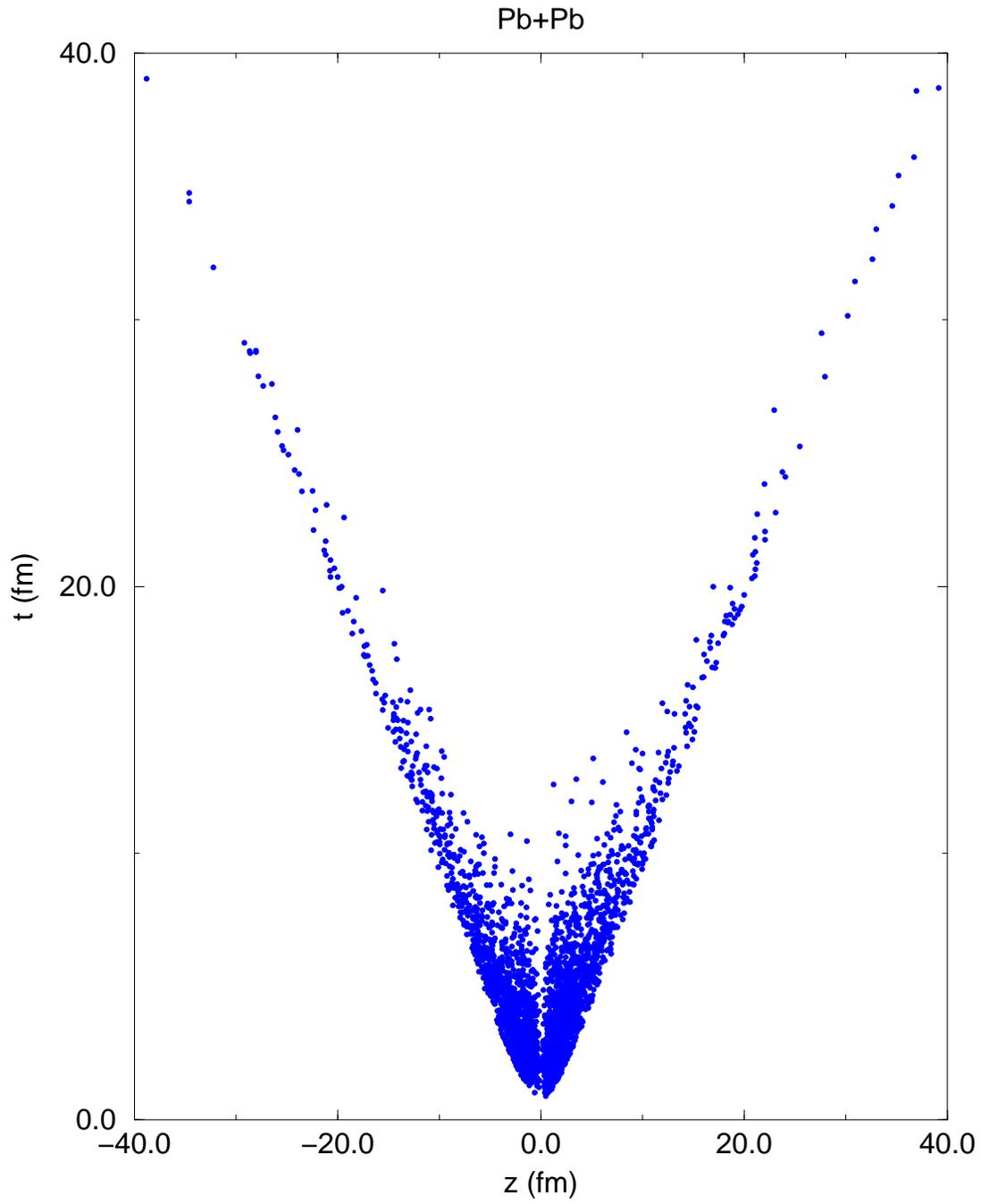}
\hfil}}
\caption[]{Space-time distribution for generic mesons in Pb+Pb after the
$\sim 1$ fm/c formation time. Mesons begin interacting in the second stage
cascade only after reaching these positions.} 
\label{fig:thirteen}
\end{figure}

The final step in the reinitialisation is to place all of the mesons and
baryons on a new initial spacelike surface, which we take to be the surface
$t=\tau_c$ in the global cascade frame. That is, the second stage, soft
cascade starts at the time of the last nucleon-nucleon collision which
occured in the initial stage. It was thought best to distribute the mesons
produced in a group randomly (that is, uniformly) along the space-time paths
followed by the baryons in that group during the initial cascade (see
Fig(\ref{fig:three})). Any sins committed in making such a somewhat arbitrary
choice are clearly and strongly remediated, since in any case a formation
time must elapse for each meson before it can actually interact. Figure
\ref{fig:thirteen} displays the mesons after the passage of the formation
time, for Pb+Pb at 158 GeV/c, in the $z$-$t$ plane. A limited variation in
the initial space-time placement of generic mesons, at the end of the fast
cascade, produces rather insignificant changes in the final spectra.  The
four-momenta, in the global frame, and final positions of the baryons are
already known from the first stage. We then restart the cascade by
constructing all possible two body collisions and decays among the meson and
baryon resonances produced in the hard cascade.

An important first approximation to keep in mind is that many of the results
of the fast cascade as described here are, for the baryons, similar to those
expected from a pure Glauber theory calculation. This point will be made more
evident in later work on $J/\psi$ production in pA and AB collisions.

\section{Low Energy Cascade} 

The final stage of the cascade involves standard two-body hadron-hadron
collisions and is carried out for the most part similarly to, for example,
the low energy cascade ARC \cite{ARC1}. However, our model of hadron-hadron
scattering and phase space, as described above, differs significantly and is
in fact necessary at the much higher energies encountered at the SPS and
RHIC.  We are employing as rescatterers the generic resonances defined
earlier, for both mesons and baryons. The cross-section tables are defined to
low enough energies, and proper thresholds are included for eventual
production of the observed low lying resonances. Only those meson
resonances likely to significantly affect the dynamics are included at this
stage, i.e. the $\pi$ and the $\rho$. Almost without exception the ensuing
collisions at SPS occur at relatively low center of mass energies,
$\sqrt{s}\sim 5-8$ GeV. The final second phase cascading can truly be
labelled as `low energy'.

Additional aspects requiring discussion involve the role of the produced
particle formation time $\tau_f$ and of course the decay time $\tau_d$ of the
generic hadrons into final state mesons. These certainly involve parameters
not easily discernible from elementary hadron-hadron measurements. Some
attempts to use HBT for determining the size of interaction regions
\cite{GGoldhaber} in these elementary collisions might, however, be viewed in
this light. The formation time is a straightforward, often used concept, and
in accordance with most estimates we take $\tau_f \sim 1$ fm/c, with some
$0.25$ fm/c attributable to the initial rapid phase at SPS energies.  The
generic resonances have rather low mass and are allowed to decay into only
two or three $\pi$'s (and/or strange mesons if strangeness is present), Given
the low values and narrow range of masses for the resonances we have, as
indicated, opted to use a single, constant decay width, $\Gamma = 125$
MeV. This choice does not affect the results in any dramatic fashion.

We intend in future to use both pA and light ion systems to finally fix the
formation and decay times. We choose to accomplish this here by focusing on
the S+S data collected by NA35\cite{NA35}, leaving more complete discussion of
pA to future work. In earlier work \cite{LUCIFERI}, the decay constant played
a significant role in normalising the pion spectrum for S+S. The extrapolation
to Pb+Pb, using of course the same decay time, then proved reasonably
successful \cite{LUCIFERI}, indicating that the treatment of energy loss in
the earlier work had some merit. This helped define one side of the fast
vs.~slow energy loss dilemma we posed at the outset. The large influence of
the decay time in the early calculation was easily understood. Lengthening
this time delayed the reinteraction of produced particles until particle
densities had decreased through expansion of the entire interacting
system. This led in turn to a lower pion yield. The reduced dependence on
these times, seen in the new double cascade can be attributed, once again, to
the lower average collision energies in the final phase.

\begin{figure}
\vbox{\hbox to\hsize{\hfil
\epsfxsize=6.1truein\epsffile[24 59 562 736]{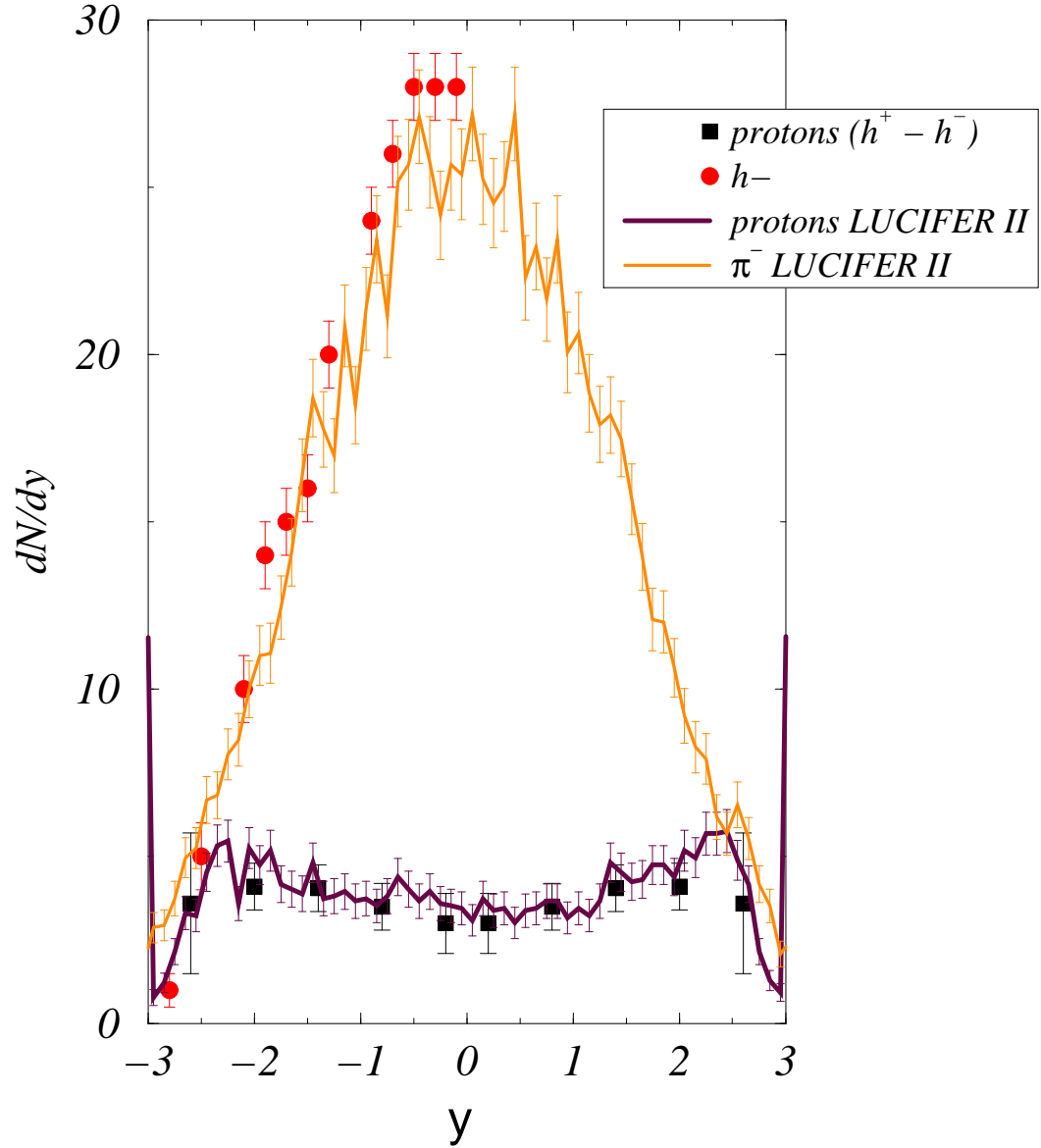}
\hfil}}
\caption[]{The calculated S+S rapidity spectra at $200$ GeV/c for $\pi^-$ and protons
compared to measurements by NA35. The latter are for total $h^-$ and  
$h^+ -h^-$ respectively, but the experimental proton equivalent spectrum was 
corrected for the $K^+$ - $K^-$ contribution.}
\label{fig:fourteen}
\end{figure}

\begin{figure}
\vbox{\hbox to\hsize{\hfil
\epsfxsize=6.1truein\epsffile[24 59 562 736]{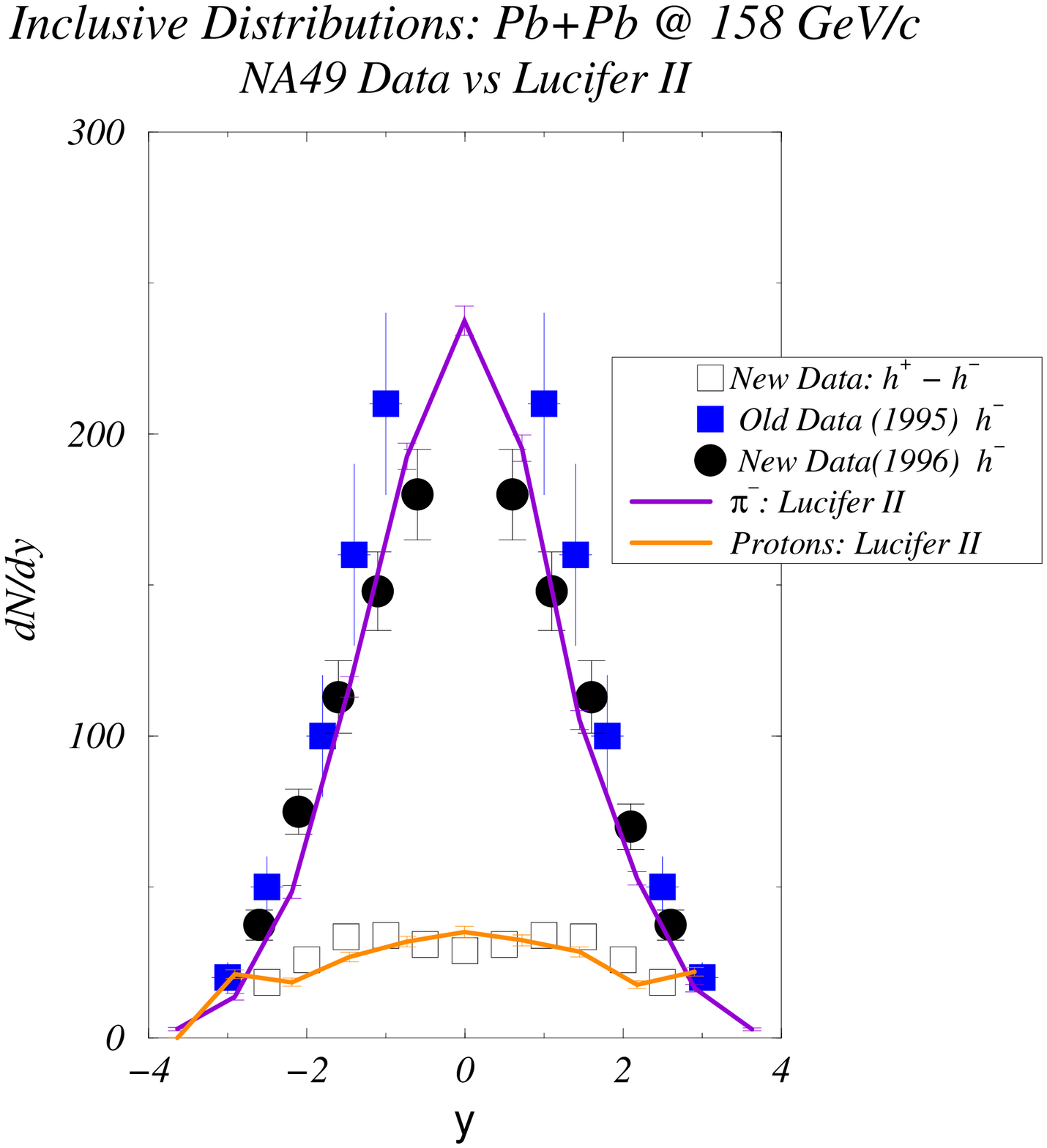}
\hfil}}
\caption[]{The calculated Pb+Pb rapidity spectra at $158$ GeV/c for $\pi^-$ and protons
compared to preliminary measurements by NA49. The latter are for total $h^-$
and $h^+ - h^-$ respectively.}
\label{fig:fifteen}
\end{figure}

The initial particle list for the soft cascade, including space-time
coordinates and four-momenta of all mesons and baryons were set in the
reinitialisation. Their rapidity distributions and the initial geometry, are
depicted in figures (4), (11), (12) and (13). A major reason for the lower
final collision energies is that rarely do particles originating separately
from the initial target and projectile configurations meet.  Spectators also
avoid each other for geometric reasons, arising as they do, from regions of
low nucleon density in the hard cascade. The results of the low energy
cascade, for a range of CERN SPS energies, are displayed in a series of
figures (14--17) for the S+S and Pb+Pb systems. These results of course
represent the output of both stages of the cascade, and wherever possible are
compared to existing data. The theoretical description of the experiments is
certainly acceptable. The proton distributions, for S+S in
Fig(\ref{fig:fourteen}) and for Pb+Pb in Fig(\ref{fig:fifteen}), measured by
the `$h^+ - h^-$ method' \cite{NA35,NA49}, show a large energy loss, perhaps
close to saturation, as suggested by the experimentalists \cite{harris} for
central collisions in the light S+S system.

The corresponding pion spectra in S+S and Pb+Pb are well reproduced by
this (Lucifer II) simulation, both in magnitude and shape. There is, however,
still a necessity for better experimental determination of both proton and
pion spectra at midrapidity in Pb+Pb at the SPS. An appreciable increase or
decrease in energy loss and meson number in the central region could have a
commensurate effect on the very interesting question of the degree of
charmonium breakup expected during the soft cascade, i.e. the amount of
$J/\psi$ suppression attributable to comovers.

Moreover these S+S and Pb+Pb results are achieved using the same parameters
for both systems; only the dependence on the number of first stage
baryon-baryon collisions plays a distinguishing role in reinitialisation.
The average number of hard collisions per nucleon in central S+S is near 3.5,
while in Pb+Pb it is closer to 7.5, with wide fluctuations about these
averages seen in both nuclear systems. Despite this difference in average
hard collision number one can begin to understand the saturation of stopping
referred to by the experimentalists \cite{harris,NA35,NA49}. At high energy
considerable energy is lost, perhaps one-half of the total, per
nucleon-nucleon collision and there is very little left to lose after just a
few collisions. Nevertheless the final soft cascading does rearrange and
broaden the rapidity distributions, even if it does not add much to the
production of a given species.

\section{Drell-Yan}

Here we present briefly the method used for calculating the other, important
side of the quandary we faced at the start: the production of massive lepton
pairs. We limit ourselves to the canonical FNAL E772 pA measurement at 800
GeV/c, but in fact the method of calculation guarantees agreement with the
lower energy pA and AA data collected by NA50\cite{NA50}. The calculation is
straightforward and is a first example, for us, of a hard process added to
the initial high energy cascade. Drell-Yan is generally considered
perturbatively calculable for dilepton pairs with masses $M_{\mu\mu}$ in
excess of $4$ GeV.  We have simply used the first Born approximation for the
production in $NN$, $N\pi$ and $\pi\pi$, with a multiplicative factor for
$p_T$ dependence and structure functions taken from early papers on the
subject\cite{NA3}. Our concern here is mainly with the overall
$A$-dependence of the process, so that including precise $Q$-factors, higher
loop corrections to the $p_T$ distributions, and using the most current
structure functions, are all unnecessary complications. We find that
contributions to Drell-Yan from meson-nucleon and meson-meson collisions,
present in the second low energy phase, are completely negligible at CERN
energies. Only the high energy nucleon-nucleon collisions in the first phase
count.

The calculation is straightforward: production in the short time
defined by such high mass pairs proceeds essentially without energy loss,
since it is extremely rare, and occurs during the initial Glauber-like hard
cascade. The simulation produces the points shown in Fig(\ref{fig:nineteen}),
which depicts the $A$-dependence of the pair yield, after division by
$A$. The E772 \cite{E772} measurements suggest a strict proportionality,
$\sigma_{DY}\sim A$. To perform the Drell-Yan microscopically we have
introduced parton structure functions\cite{NA3} and the requisite parton
variables, into the coding. But the theoretical results in
Fig(\ref{fig:nineteen}) could have been obtained purely geometrically from
the elementary production rates and from the space-time history of the first,
high energy, nucleon-nucleon cascade.

\section{Further Results and Conclusions}

The main purpose of this manuscript has been to present the theoretical
underpinnings of a modified cascade simulation of high energy heavy ion
collisions that can reasonably describe two seemingly contradictory aspects
of the data. The observed $A$-dependence of cross-sections for high mass
Drell-Yan production in proton-nucleus and ion-ion collisions at SPS energies
and above, suggests that high mass lepton pairs are produced by nucleons
which {\it do not experience energy loss} during the ion-ion collision, while
the observed meson and baryon spectra make it all too clear that energy is
indeed lost from the baryons and goes into producing soft mesons. A standard
two-body cascade model cannot easily reconcile these facts, since energy is
necessarily lost with each successive two-body collision. We have suggested a
solution to this quandary: the simulation should be done in three
steps. First a short-time high energy Glauber-like cascade is carried out
among the nucleons involving only energy loss resulting from
high momentum transfer processes. This builds in the basic space-time
geometry of the ion-ion collision. Second the cascade is reinitialised,
allowing for soft meson production from the nucleons. This production is done
coherently, from groups of interacting nucleons, but two-body inputs are used
together with the collision history to strongly constrain the meson
production. Finally, an ordinary low energy two-body cascade is carried out
among all the particles produced in the first two stages.  This modified
cascade approach becomes essential at collision energies near and above those
attained at the SPS, where the total time for the nuclei to pass through
eachother (in an equal velocity frame) becomes comparable to the time for
soft meson formation ($\tau_c \sim \tau_f \sim 1$ fm).

\begin{figure}
\vbox{\hbox to\hsize{\hfil
\epsfxsize=6.1truein\epsffile[24 59 562 736]{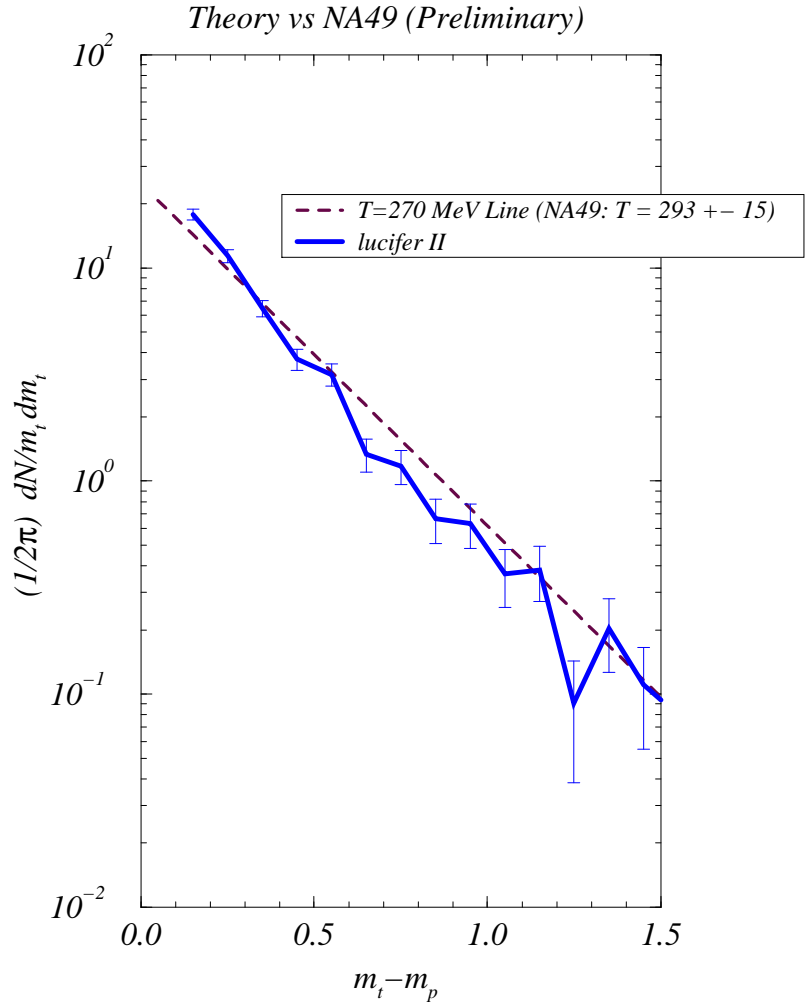}
\hfil}}
\caption[]{A calculated transverse momentum spectrum for protons in Pb+Pb
using a rapidity interval $-1.0\le y \le +1.0]$. Comparison is made to an exponential
with inverse slope or `temperature' T close to
270 MeV. The preliminary central NA49 value is is T=$293 \pm 10$ MeV, for
$-0.25\le y\le 0.25]$.}
\label{fig:sixteen}
\end{figure}

We did not present here an inclusive treatment of all existing data, but
limited ourselves to a few applications selected from CERN SPS
\cite{NA35,NA49}, UA5, FNAL \cite{Eisenberg,E772} experiments, to demonstrate
that the ideas can work in practice. Some of these examples are from pA
collisions and some from AA. In addition to the rapidity distributions for
S+S and Pb+Pb, we have also considered transverse momentum distributions;
examples are shown in Fig(\ref{fig:sixteen}) for protons from Pb+Pb in
central rapidity slices and also for $\Lambda$'s in comparison to recent
preliminary data from Quark Matter '96\cite{NA49}. The calculated and
measured `temperatures' are in reasonably good agreement.

Nevertheless, the degree of proton stopping \cite{harris} is clearly
important to our model and a better understanding of the data might point to
differences between theory and experiment, heralding collective effects. It
will be particularly important to see the differences occasioned by the
introduction of partons. Results of parton cascades such as that of K.~Geiger
\cite{Geiger1}, when compared to our preliminary calculations for central
Au+Au collisions at RHIC energy, suggest remarkably similar spectra may
result for the mesons when partonic hard scattering and splitting processes
are included. This perhaps suggests that two-body processes and energy
conservation, not thermodynamics, dominate such production. Despite this, the
energy and number densities for both baryons and mesons achieved in the
present simulations (LUCIFER II) are high enough, $\rho_E \sim 3-5$
GeV/fm$^3$, and last for sufficiently long times in central Pb+Pb collisions
at 158 GeV/c, perhaps to expect unusual high density behaviour to occur. The
striking new results of {NA50} \cite{NA50} must then be taken very seriously.

We have also considered the leading particle behaviour for pp and p+Pb, and
illustrate these in a comparison with 100 GeV/c FNAL measurements        
\cite{earlypAdata} of proton and pion spectra in Feynman $x$ in
Fig(\ref{fig:eighteen}). A key feature in understanding this low $p_T$ data
is the narrower transverse momentum present in both elastic and inelastic
components in the $NN$ model for Feynman $x$ close to $1$. Although these 
results are perhaps related to the parton structure of the proton, a simple
conclusion is that the nucleon-nucleus data follows from nucleon-nucleon.

The very interesting case of charmonium production in both pA \cite{E772} and
AA \cite{NA50,Kharzeev,gavin}, should receive some mention. We have already
treated the pA charmonium production in a somewhat different
fashion\cite{LUCIFERI}, following suggestions by Gottfried\cite{gottfried}.
The model we used\cite{LUCIFERI} treated the charmonium states as coupled
channels, at least to the extent that the final $\chi$ states, or some
pre-resonant form of them \cite{Kharzeev}, are much more abundantly produced
than the $J/\psi$.  Certainly the $\chi$ states feed strongly into the $\psi$
by electromagnetic decay. The parameters in such a model are the breakup
cross-sections of the charmonium states in the initial pure nucleon
cascade. The very similar suppression in pA found for $\psi'$ and $\psi$ by
the FNAL E772 experiment \cite{E772} can perhaps be understood if one takes
the coupled channels approach seriously and ties the breakup probability to
the size of the charmonium state, since both the $\chi$ and $\psi'$ are
considerably larger than the $J/\psi$. In any case we will separately
present a calculation of charmonium yields based on the current modelling.

It has of course been the principal thrust of this work to create a unified
dynamic approach to calculating ion-ion collisions at very high energy,
incorporating both hard processes such as Drell-Yan and the slower processes
responsible for energy loss and soft meson production. So far, this
has been done allowing rescattering only via hadronic intermediate states. No
partons have yet been explicitly included, except insofar as structure
functions must be used for the calculation of Drell-Yan production. This
approach proves to be phenomenologically viable and leads to a reasonable
description of a broad range of results obtained at the SPS. A secondary
justification for constructing such a pure hadronic calculation is to
investigate the conclusions of Kharzeev and Satz \cite{Kharzeev}, namely that
a purely hadronic explanation of the $J/\psi$ suppression seen by NA50 is not
possible.

\begin{acknowledgements}

The authors are of course grateful to all of the experimentalists at
the AGS, SPS, ISR and FNAL, who have over the years gathered the basic
data. Certainly the code could not have been constructed without it. We also
wish to thank Y.~Dokshitzer, A.~H.~Mueller for instructive
conversations and advice. This manuscript has been authored under US DOE
contracts No. DE-FG02-93ER407688 and DE-AC02-76CH00016.
\end{acknowledgements}

\vfill\eject

\clearpage

\clearpage

\clearpage

\clearpage

\clearpage

\clearpage

\clearpage

\clearpage

\clearpage

\clearpage

\clearpage

\clearpage

\clearpage

\clearpage

\clearpage

\clearpage

\clearpage

\begin{figure}
\vbox{\hbox to\hsize{\hfil
\epsfxsize=6.1truein\epsffile[24 59 562 736]{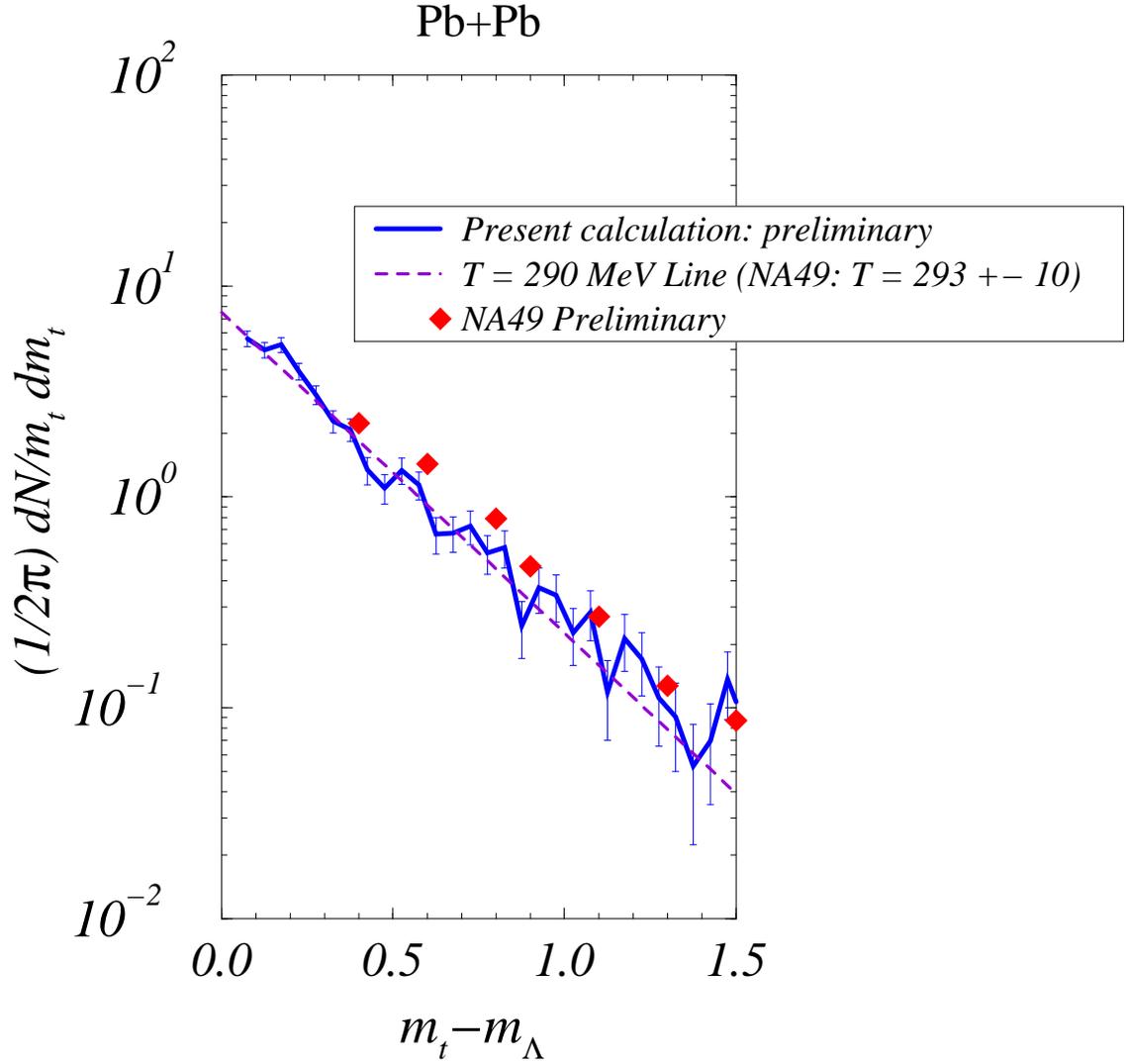}
\hfil}}
\caption[]{A calculated transverse momentum spectrum for $\Lambda$ in Pb+Pb
using a rapidity interval $-1.0\le y\le +1.0$. Comparison is made to
preliminary NA49 data for central Pb+Pb.}
\label{fig:seventeen}
\end{figure}
\clearpage

\begin{figure}
\vbox{\hbox to\hsize{\hfil
\epsfxsize=6.1truein\epsffile[24 59 562 736]{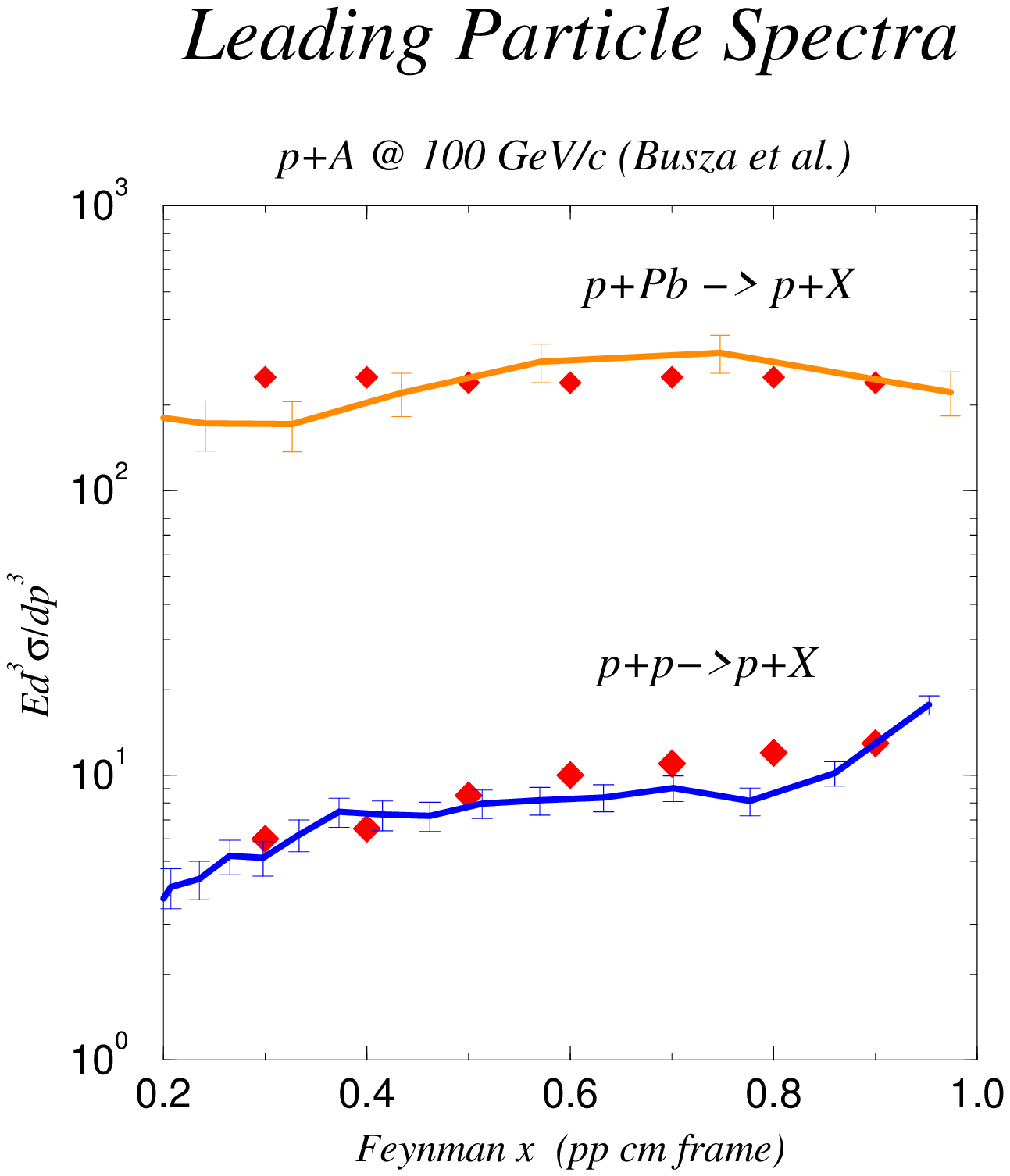}
\hfil}}
\caption[]{Leading particle behaviour for inclusive proton production
in pp and p+Pb. Simulations at 100 GeV/c are compared as a function of Feynman
x to FNAL data from Busza et al.\cite{earlypAdata}.}
\label{fig:eighteen}
\end{figure}
\clearpage

\begin{figure}
\vbox{\hbox to\hsize{\hfil
\epsfxsize=6.1truein\epsffile[24 59 562 736]{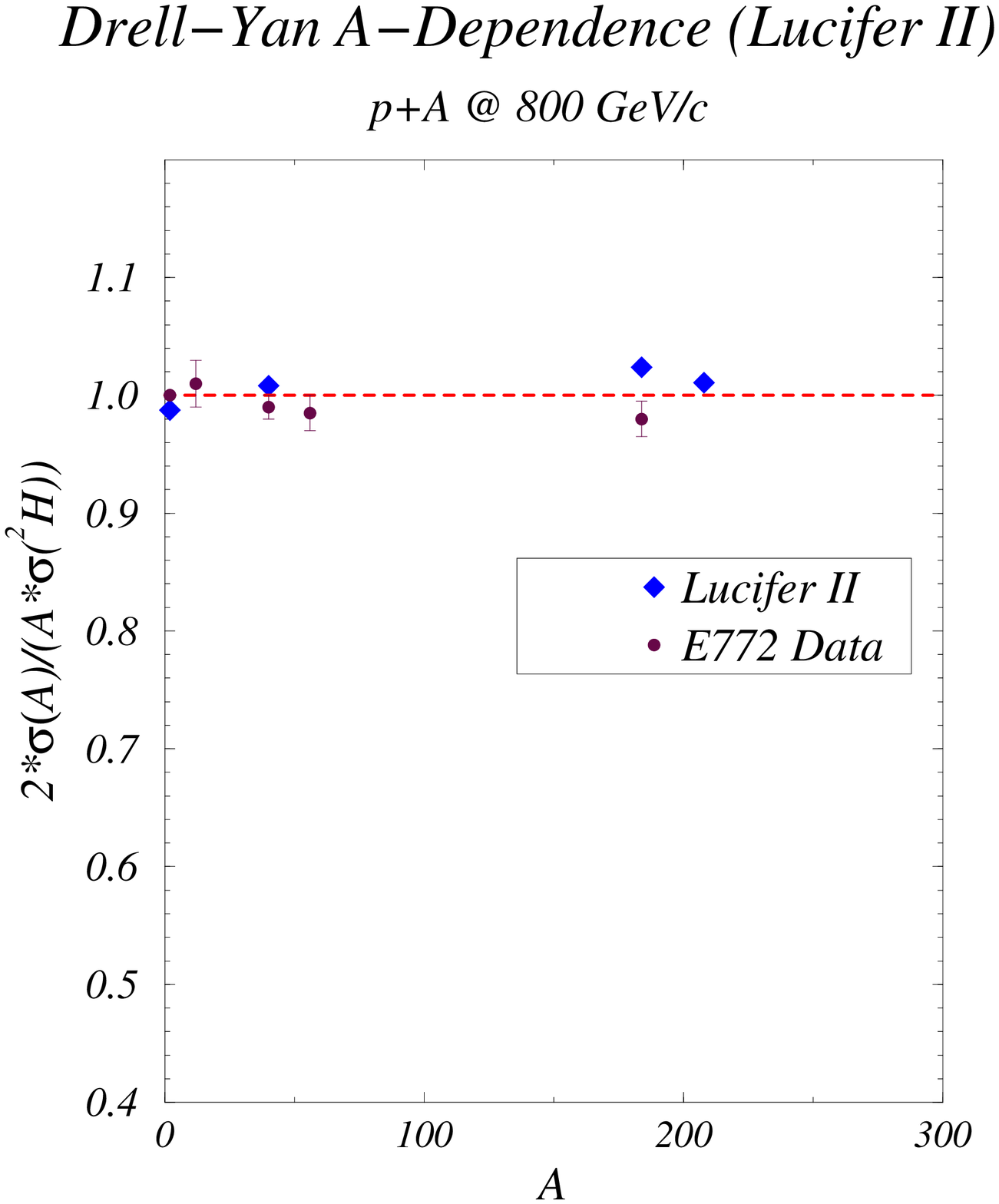}
\hfil}}
\caption[]{Drell-Yan for pA. Similar to Figure 1 but with theoretical
points added for a selection of nuclei. The only appreciable calculated
production comes from the high energy phase and is essentially geometrical.}
\label{fig:nineteen}
\end{figure}
\clearpage

\end{document}